\pdfoutput=1 
\documentclass[journal]{IEEEtran}
\ifCLASSINFOpdf
\else
\fi
\usepackage{hyperref}
\usepackage{svg} 
\usepackage{xcolor,soul} 
\usepackage{amsmath,graphicx,amsfonts}

\begin{document}
%
\title{Multi-channel Multi-frame ADL-MVDR for Target Speech Separation}

\author{Zhuohuang Zhang\thanks{This work was done while Z. Zhang was an intern at Tencent AI Lab, Bellevue, USA.}, \textit{Student Member}, \textit{IEEE}, Yong Xu, Meng Yu, Shi-Xiong Zhang, Lianwu Chen, \\
Donald S. Williamson, \textit{Member}, \textit{IEEE}, Dong Yu, \textit{Fellow}, \textit{IEEE}  
\thanks{Zhuohuang Zhang is with the Department
of Computer Science and Department of Speech, Language and Hearing Sciences, Indiana University, Bloomington,
IN, 47408 USA (e-mail: \href{mailto: zhuozhan@iu.edu}{zhuozhan@iu.edu}).}
\thanks{Yong Xu, Meng Yu, Shi-Xiong Zhang, and Dong Yu are with Tencent AI Lab, Bellevue, WA 98004 USA (e-mail: \href{mailto: lucayongxu@tencent.com}{lucayongxu@tencent.com}; \href{mailto: raymondmyu@tencent.com}{raymondmyu@tencent.com}; \href{mailto: auszhang@tencent.com}{auszhang@tencent.com};  \href{mailto: dyu@tencent.com}{dyu@tencent.com}).}
\thanks{This work was done when Lianwu Chen was a researcher in Tencent AI Lab, Shenzhen, China.}
\thanks{Donald S. Williamson is with the Department
of Computer Science, Indiana University, Bloomington,
IN, 47408 USA (e-mail: \href{mailto: williads@indiana.edu}{williads@indiana.edu}).}
}

\markboth{}%
{Shell \MakeLowercase{\textit{et al.}}: Bare Demo of IEEEtran.cls for IEEE Journals}

\maketitle

\begin{abstract}
Many purely neural network based speech separation approaches have been proposed to improve objective assessment scores, but they often introduce nonlinear distortions that are harmful to modern automatic speech recognition (ASR) systems. Minimum variance distortionless response (MVDR) filters are often adopted to remove nonlinear distortions, however, conventional neural mask-based MVDR systems still result in relatively high levels of residual noise. Moreover, the matrix inverse involved in the MVDR solution is sometimes numerically unstable during joint training with neural networks. In this study, we propose a multi-channel multi-frame (MCMF) all deep learning (ADL)-MVDR approach for target speech separation, which extends our preliminary multi-channel ADL-MVDR approach. The proposed MCMF ADL-MVDR system addresses linear and nonlinear distortions. Spatio-temporal cross correlations are also fully utilized in the proposed approach. The proposed systems are evaluated using a Mandarin audio-visual corpus and are compared with several state-of-the-art approaches. Experimental results demonstrate the superiority of our proposed systems under different scenarios and across several objective evaluation metrics, including ASR performance.

\end{abstract}


\IEEEpeerreviewmaketitle

\section{Introduction}
\IEEEPARstart{T}{arget} speech separation algorithms extract target speech signal from noisy background when interfering sources and background noise exist~\cite{wang2018supervised}. These algorithms serve as important front-ends for many speech communication systems such as automatic speech recognition (ASR) \cite{du2014robust,weninger2015speech,zhang2017speech,AVJW2020}, speaker verification \cite{eskimez2018front}, and digital hearing-aid devices \cite{van2009speech}. With the recent achievements in deep learning, many neural network (NN) based speech separation systems have been proposed. Many early approaches synthesize the separated speech after combining the time-frequency (T-F) masked spectrogram with the original noisy phase \cite{wang2014training,erdogan2016improved,zhang2020loss}. The use of the noisy phase sets a sub-optimal upper bound on the system's performance as phase plays an important role in the perceptual speech quality and intelligibility \cite{paliwal2011importance, williamson2016complex, xu2017distorting, zhang2020investigation}. Phase-aware T-F masks have later been proposed, including the phase-sensitive mask \cite{erdogan2015phase,lee2018phase,zhang2020loss}, and complex ratio mask \cite{williamson2016complex}. Yet an accurate estimate of the phase component is still difficult for an NN to learn, due to the lack of structure in the phase response. 

Besides T-F mask based systems, many recent speech separation systems have been proposed that operate directly on the time-domain speech signal in an end-to-end fashion \cite{pascual2017segan,wang2018end,luo2018tasnet,fu2018end,stoller2018wave,luo2019conv}, to avoid directly estimating the magnitude and phase components. Some of these approaches (e.g., Wave-U-Net~\cite{stoller2018wave} and TasNet~\cite{luo2018tasnet}) replace the conventional STFT and inverse STFT (iSTFT) signal processing procedures with a learnable NN-based encoder and decoder structure. The encoded features are then altered by a learned-latent mask, where they are later fed to the decoder. The recent time-domain fully-convolutional Conv-TasNet \cite{luo2019conv} has substantially improved performance according to many objective measures, where it features a TasNet-like encoder-decoder structure that extracts the target speech in a learned latent space \cite{luo2018tasnet}. Alternatively, other approaches implicitly combine the feature extraction and the separation steps as reported in \cite{pascual2017segan,fu2018end}. 

Purely NN-based speech separation systems have achieved impressive objective speech quality scores, since they greatly reduce the amount of noise or interfering speech. These approaches, however, often introduce unwanted nonlinear distortions into the separated signal, since these models focus on removing unwanted interferences without imposed constraints that limit the solution space. These resulting nonlinear distortions negatively affect the performance of ASR systems \cite{xu2020neural,luo2019conv, tan2020audio}. To alleviate the nonlinear distortion issue and achieve better performance on source separation, speech separation systems in practice often adopt a multi-channel processing scheme to further leverage spatial information in addition to original spectral information. Many approaches have been developed, including the multi-channel Wiener filter (MWF) \cite{huang2008analysis,souden2009new}, the linearly constrained minimum variance (LCMV) filter~\cite{van1988beamforming} and the minimum variance distortionless response (MVDR) filter. The MVDR filter can be viewed as a special case of MWF and LCMV, as it forces a distortionless response when oracle directional information is available~\cite{gannot2001signal, benesty2008microphone, souden2010study, pan2013performance}. MVDR filters have been widely used in speech separation systems to reduce the amount of nonlinear distortions, which is helpful to ASR systems \cite{higuchi2016robust, xiao2017time, xu2020neural}. The distortionlessness of the separated speech is ensured as the MVDR filter is derived under constraints that preserve speech information at the target direction. On the contrary, other beamformers such as the Generalized Eigenvalue (GEV) beamformer \cite{warsitz2007blind,heymann2015blstm} aim to improve the signal-to-noise ratio (SNR) without controlling the amount of distortions in the separated speech signal. Additionally, multi-frame MVDR (MF-MVDR) filters~\cite{huang2011multi, schasse2014estimation, fischer2018robust} have been adopted in single-channel speech separation systems to remove the noise and ensure the distortionlessness of the separated speech. Prior studies have shown that when oracle information is available, the MF-MVDR filter can greatly diminish the noise while introducing few distortions~\cite{huang2011multi,fischer2017sensitivity}. 

Recent MVDR approaches are often combined with an NN-based T-F mask estimator \cite{xiao2017time,xu2019joint,tammen2019dnn,xu2020neural} that leads to more accurate estimates of the speech and noise components, and better ASR performance due to fewer nonlinear distortions. However, many of these conventional neural mask-based MVDR systems result in high levels of residual noise (e.g., linear distortions), since segment- or utterance-level beamforming weights cannot fully eliminate residual noise on each frame~\cite{habets2013two,erdogan2016improved,xiao2017time,xu2020neural}. Our recent work further incorporates multi-frame (MF) information during beamforming weights derivation \cite{xu2020neural}, where it exploits extra inter-frame correlation in addition to the spatial correlation between the microphones in conventional multi-channel MVDR approaches. Results showed better ASR accuracy and higher PESQ scores when compared to conventional neural mask-based MVDR approaches. Unfortunately, the amount of residual noise in the separated signal is still high.

Many online beamforming weights estimation approaches have been proposed recently for real-time or time-varying purposes~\cite{souden2011integrated, taseska2017nonstationary, higuchi2017online, higuchi2018frame, chakrabarty2019time, martin2020online}. In~\cite{souden2011integrated}, Souden et al.  proposed a recursive method with heuristic updating factors to estimate the time-varying speech and noise covariance matrices, but these heuristic updating factors are hard to determine and often limit the system's performance. Systems such as \cite{taseska2017nonstationary, chakrabarty2019time, martin2020online} also used smoothing factors to estimate the time-varying covariance matrices, although these approaches allow better performance in time-varying conditions, the residual noise could still be present and postfilering is sometimes needed. \cite{higuchi2018frame} proposed a frame-level beamforming method, however, it achieved slightly worse ASR accuracy compared to conventional segment-level beamforming systems \cite{higuchi2018frame}.

In the current study, we propose a novel all deep learning MVDR (ADL-MVDR) framework that can be adapted for speech separation under different microphone configurations, including multi-channel (single-frame), multi-frame (i.e., when only one channel is available to the beamforming module), and multi-channel multi-frame (MCMF) scenarios. This study extends our preliminary work on the ADL-MVDR beamformer~\cite{zhang2021adl}, which has proven to work well on multi-channel (MC) speech separation tasks. The ADL-MVDR beamformer incorporates a front-end complex filter estimator (i.e., a Conv-TasNet variant based on our prior work \cite{tan2020audio,gu2020multi}) that consists of dilated 1-D convolution blocks for speech and noise component estimation and another ADL-MVDR module for frame-level MVDR beamforming weights estimation. In contrast to conventional per T-F bin mask-based approaches, complex ratio filtering (denoted as cRF) \cite{mack2019deep} is used for more accurate estimates of the speech and noise components, while also addressing issues with handling phase. Earlier approaches have verified the idea of applying NNs for matrix inverse \cite{oja1982simplified,wang1993recurrent,fyfe1997neural, zhang2005design} and principal component analysis (PCA) \cite{oja1992principal,fyfe1997neural}. The proposed ADL-MVDR module deploys two recurrent neural networks (RNNs) to replace the matrix operations (i.e., matrix inverse and principal eigenvector extraction) performed on the noise and speech covariance matrices, respectively. Leveraging on the temporal properties of RNNs, the statistical variables (i.e., inverse of noise covariance matrix and steering vector) are estimated adaptively at the frame-level, enabling the derivation of time-varying beamforming weights, which is more suitable for diminishing the non-stationary noise at each frame. The system also uses visual information (described in our prior work~\cite{tan2020audio,gu2020multi}) to extract the direction of arrival (DOA) of the target speaker. Results from our prior study \cite{zhang2021adl} indicate that for MC speech separation tasks, the ADL-MVDR system can greatly suppress the residual noise while also ensuring that fewer distortions are introduced into the separated speech signal when compared to conventional neural mask-based MVDR approaches. 

The major contributions of this work consist of the following. Firstly, we verify the idea of applying the ADL-MVDR framework to MF speech separation tasks, when only one channel of the signal is available to the beamforming module, to further evaluate generalization. Secondly, we further adapt the ADL-MVDR framework to an MCMF speech separation task for spatio-temporal speech separation, which has not been previously done, to determine if additional MF information leads to further improvements. Thirdly, we examine and quantify the influence of the cRF and MF sizes on the performance of different ADL-MVDR systems.

The rest of this paper is organized as follows. Section \ref{sec:Signal Model} describes the signal models for conventional neural mask-based MC and MF-MVDR systems. The proposed ADL-MVDR system is revealed in Section \ref{sec:Proposed ADL-MVDR Beamformer}. The experimental setup is given in Section \ref{sec:Experimental Setup}. We present and discuss the results in Section \ref{sec:Results and Analysis}. Finally, we conclude our work in Section \ref{sec:Conclusions}.

\section{Conventional neural Mask-based MVDR Filter}
\label{sec:Signal Model}
In this section, we discuss the signal models of conventional neural mask-based MVDR filters under two different conditions, i.e., MC and MF-MVDR speech separation.

\subsection{Conventional neural Mask-based Multi-channel MVDR}
\label{subsec:Multi-channel Speech Separation}

In the MC speech separation scenario, consider a time-domain noisy speech signal $\mathbf{y} = [y^{(0)}, y^{(1)}, ..., y^{(M-1)}]^T$ recorded by an $M$-channel microphone array, where $y^{(i)}$ is the signal recorded from the $i$-th channel. Let $\mathbf{Y}(t,f)$, $\mathbf{X}(t,f)$ and $\mathbf{N}(t,f)$ denote the $M$-dimensional T-F domain MC noisy-reverberant speech, reverberant speech and noise signals, respectively. We have 
\begin{equation}
\mathbf{Y}(t,f) = \mathbf{X}(t,f) + \mathbf{N}(t,f), 
\end{equation}
where $(t,f)$ represents the corresponding frame and frequency indices. Note that we use reverberant speech as the learning target as we focus on separation only in this study. In the time-domain, we have $\mathbf{x}(n) = \mathbf{g}(n)*\mathbf{s}(n)$, where $\mathbf{x}(n)$ and $\mathbf{s}(n)$ are the time-domain MC reverberant and anechoic speech signals, $n$ is the time index, $\mathbf{g}(n)$ represents the room impulse response and `$*$' denotes linear convolution. 

The estimated single-channel reverberant speech, $\mathrm{\hat{X}^{(0)}}(t,f)$, can be obtained with the MC-MVDR filter as
\begin{equation}
\label{eq:beamforming}
\mathrm{\hat{X}^{(0)}}(t,f) = \mathbf{h_{\text{MC-MVDR}}^{\mathrm{H}}}(f)\mathbf{Y}(t,f),
\end{equation}
where $\mathbf{h_\text{MC-MVDR}}(f) \in \mathbb{C}^{M}$ denotes the MC-MVDR beamforming weights, and $^{\mathrm{H}}$ is the Hermitian operator. The objective of MC-MVDR filtering is to minimize the power of the noise without introducing distortions into the target speech signal. This can be formulated as
\begin{equation}
\label{eq:mc-mvdr-constraint}
\mathbf{h_{\text{MC-MVDR}}}=\underset{\mathbf{h}}{\arg \min \mathbf{h}}^{\mathrm{H}} \mathbf{\Phi}_{\mathbf{NN}} \mathbf{h} \quad \bf{\text {s.t.}} \quad \mathbf{h}^{\mathrm{H}} \mathbf{\boldsymbol{v}}=\mathbf{1},
\end{equation}
where $\boldsymbol{v}(f) \in \mathbb{C}^{M}$ stands for the target speech steering vector, which can be estimated as the principal eigenvector of speech covariance matrix (i.e., $\boldsymbol{v}(f) = \mathcal{P}\{\mathbf{\Phi}_{\mathbf{XX}}\}$) \cite{higuchi2016robust,heymann2016neural,liu2018neural}. $\mathbf{\Phi}_{\mathbf{NN}}$ represents the noise covariance matrix. The MVDR solution based on the steering vector is \cite{shimada2018unsupervised,wang2018all}
\begin{equation}
\label{eq:MVDR_solutions}
\begin{aligned}
\mathbf{h_{\text{MC-MVDR}}}(f)&=\frac{\mathbf{\Phi}_{\mathbf{NN}}^{-1}(f) \boldsymbol{v}(f)}{\mathbf{\boldsymbol{v}^{\mathrm{H}}}(f) \mathbf{\Phi}_{\mathbf{NN}}^{-1}(f) \mathbf{\boldsymbol{v}}(f)}.
\end{aligned}
\end{equation}
In practice, many studies adopt the other MVDR solution that is based on the reference channel \cite{habets2013two,xu2020neural,subramanian2019investigation}
\begin{equation}
\label{eq:MVDR2_solutions}
\begin{aligned}
\mathbf{h_{\text{MC-MVDR}}}(f)&=\frac{\mathbf{\Phi}_{\mathbf{NN}}^{-1}(f) \mathbf{\Phi}_{\mathbf{XX}}(f)}{\text{Trace}(\mathbf{\Phi}_{\mathbf{NN}}^{-1}(f) \mathbf{\Phi}_{\mathbf{XX}}(f))}\mathbf{u},
\end{aligned}
\end{equation}
where $\mathbf{u}$ is the one-hot vector that selects the reference microphone channel. Note that the matrix inverse involved in the MVDR solution is sometimes not numerically stable during joint training with NNs. For instance, the estimated noise covariance matrix could be singular, which will cause numerical instability when computing the matrix inverse~\cite{zhao2012fast, lim2017numerical, zhang2021end}, where diagonal loading is often used to alleviate this issue~\cite{mestre2003diagonal,xu2019joint,xu2020neural, zhang2021end}.

In a typical neural mask-based MVDR system, the covariance matrices are estimated chunk-wisely with T-F masks \cite{heymann2016neural, erdogan2016improved,tammen2019dnn,wang2018spatial,xu2020neural}. A system that uses a real-valued T-F mask (RM) (e.g., with an ideal binary mask (IBM), ideal ratio mask (IRM), ...) performs covariance matrix estimation as follows
\begin{equation}
\begin{aligned}
\label{eq:tf_mask_cov_mat}
\mathbf{\hat{\Phi}}_{\mathbf{XX}}(f) &= \frac{\sum_{t=1}^{T} \mathrm{RM}^{2}_{\mathrm{X}}(t,f) \mathbf{Y}(t,f) \mathbf{Y}^{\mathrm{H}}(t,f)}{\sum_{t=1}^{T} \mathrm{RM}^{2}_{\mathrm{X}}(t,f)},
\end{aligned}
\end{equation}
where $\mathrm{RM}_{\mathrm{X}}$ stands for the RM for estimating the speech component and $T$ is the total number of frames. The power mask is used for normalization. The noise covariance matrix $\mathbf{\hat{\Phi}}_{\mathbf{NN}}$ can be computed in a similar manner. Nevertheless, we want to point out that the chunk-wise beamforming cannot fully eliminate residual noise on each frame and therefore the relatively high levels of residual noise becomes a hand-in-hand problem for conventional neural mask-based MC-MVDR systems.

\subsection{Conventional Multi-frame MVDR}
\label{subsec:Multi-frame Speech Separation}

\begin{figure*}[t!]
  \centering
  \includegraphics[scale = 0.47]{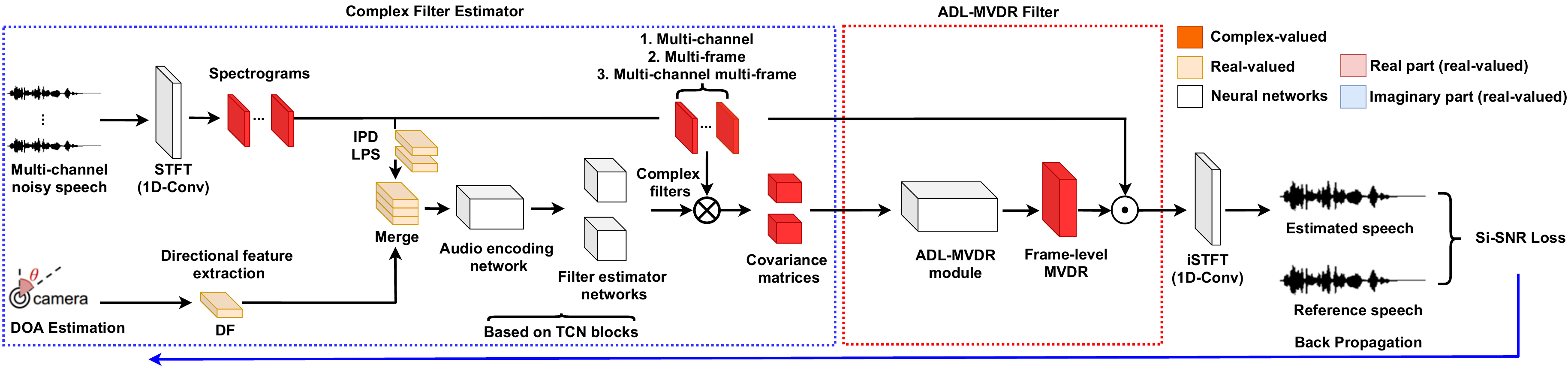}
  \caption{(Color Online). Network architecture of our proposed ADL-MVDR framework. It consists of a complex filter estimator (i.e., highlighted in blue dashed box) based on temporal convolution network (TCN) blocks for components estimation (depending on the case), and an ADL-MVDR filter (i.e., highlighted in red dashed box) that consists of two GRU-Nets for frame-wise MVDR coefficients estimation. $\otimes$ and $\odot$ denote the operations expressed in Eq. (\ref{eq:cRF_filtering}) and (\ref{eq:tf-beamforming}), (\ref{eq:sc_mvdr_solution}) or (\ref{eq:tf-mcmfbeamforming}), respectively, depending on the situation.}
  \label{fig:ADL-MVDR}
\end{figure*}

\begin{figure}[t]
  \centering
  \includegraphics[scale = 0.82]{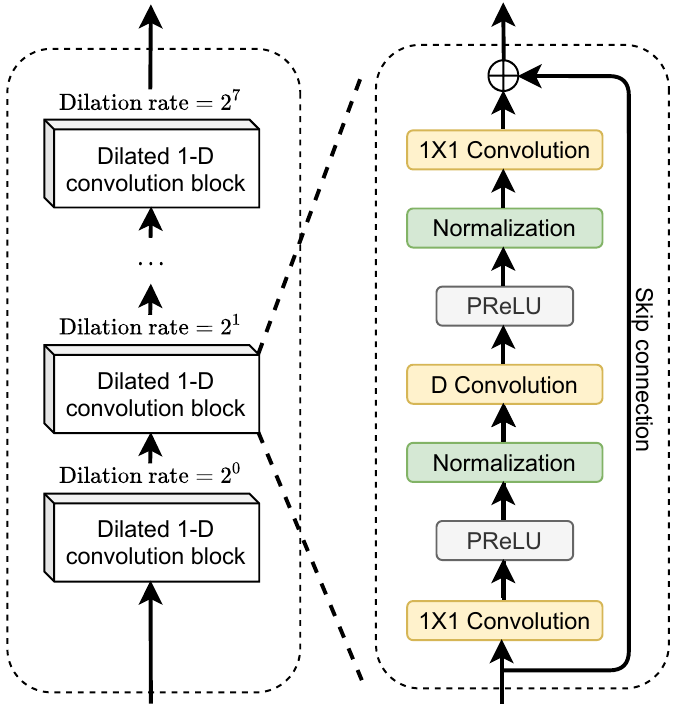}
  \caption{(Color Online). Illustration of one TCN block (left) that consists of 8 dilated 1-D convolution blocks (right). Each dilated 1-D convolution block consists of 1$\times$1 (i.e., pointwise) convolution layers and a depth-wise separable convolution layer (D convolution)~\cite{8099678}. PReLU~\cite{he2015delving} activation and normalization are applied between the convolution layers. Skip connection is also used.}
  \label{fig:audioencodingblock}
\end{figure}

The MF-MVDR filter can be viewed as a special extension of the MC-MVDR beamformer when only one channel of the signal is available. In this case, the spatial information is lost and therefore, the MF-MVDR filter tries to explore the interframe correlations instead of performing spatial beamforming. Many MF-MVDR systems have been proposed recently \cite{huang2011multi,benesty2011single, tammen2019dnn}. Analogous to the MC-MVDR speech separation scenario, the process of obtaining the MF-MVDR enhanced speech can be described as
\begin{equation}
\label{eq:sc-beamforming}
\mathrm{\hat{X}^{(0)}}(t,f) = \mathbf{\mathbf{h^{\mathrm{H}}_{\text{MF-MVDR}}}}(t,f)\mathbf{\overline{Y}^{(0)}}(t,f),
\end{equation}
where $\mathbf{h_{\text{MF-MVDR}}}(t,f) \in \mathbb{C}^{L}$ and $\mathbf{\overline{Y}^{(0)}}$ represent the L-dimensional MF-MVDR filter coefficients and L consecutive STFT frames of the single-channel noisy speech signal~\cite{huang2011multi,benesty2011single,xu2020neural}, respectively,
\begin{equation}
\begin{aligned}
\label{eq:sc-h and y}
\mathbf{h_{\text{MF-MVDR}}}(t,f) &= [\mathrm{h}_{0}(t,f),\mathrm{h}_{1}(t,f),...,\mathrm{h}_{L-1}(t,f)]^{T}, \\
\mathbf{\overline{Y}^{(0)}}(t,f) &= [\mathrm{Y^{(0)}}(t,f),\mathrm{Y^{(0)}}(t-1,f),...,\\
&\mathrm{Y^{(0)}}(t-L+1,f)]^{T},
\end{aligned}
\end{equation}
where $h_{l}(t,f)$ represents the $l$-th filter coefficient and $\mathrm{Y^{(0)}}(t,f)$ is the single-channel noisy speech STFT. This is similar to MC beamforming methods by viewing different frames as microphone inputs of different channels. Note that Eq. (\ref{eq:sc-h and y}) can be extended to use information from future frames, which benefits speech processing. The MF speech and noise can be constructed by concatenating the delayed (or shifted for future frames) version of estimated speech or noise components following similar steps in Eq. (\ref{eq:sc-h and y}). The objective of the MF-MVDR filter is also to minimize the power of the interfering sources while preserving the components from the target speech, which can be computed as
\begin{equation}
\label{eq:sc_mvdr}
\mathbf{h_{\text{MF-MVDR}}}=\underset{\mathbf{\overline{h}}}{\arg \min \mathbf{\overline{h}}}^{\mathrm{H}} \mathbf{\Phi}^{\text{MF}}_{\mathbf{VV}} \mathbf{\overline{h}} \quad \bf{\text {s.t.}} \quad \mathbf{\overline{h}}^{\mathrm{H}} \mathbf{\boldsymbol{\gamma}_{x}}=\mathbf{1},
\end{equation}
where $\mathbf{\Phi}^{\text{MF}}_{\mathbf{VV}}(t,f) \in \mathbb{C}^{L \times L}$ denotes the covariance matrix of the MF undesired signal component~\cite{huang2011multi,schasse2014estimation,tammen2019dnn} which consists of the noise and the uncorrelated speech components. $\mathbf{\boldsymbol{\gamma}_{x}}(t,f) \in \mathbb{C}^{L}$ is the speech interframe correlation (IFC) vector that describes the correlation between the previous and current frames. According to \cite{benesty2011single,tammen2019dnn}, the speech IFC vector $\mathbf{\boldsymbol{\gamma}_{x}}$ can be formulated as
\begin{equation}
\begin{aligned}
\mathbf{\boldsymbol{\gamma}_{x}}(t,f)=\frac{\boldsymbol{\Phi}^{\text{MF}}_{\mathbf{XX}}(t,f) \mathbf{e}}{E[|\mathrm{X}^{(0)}(t,f)|^2]},
\end{aligned}
\end{equation}
where $\mathbf{\Phi}^{\text{MF}}_{\mathbf{XX}}$ stands for the covariance matrix of the MF speech and $\mathrm{X}^{(0)}$ represents the single-channel speech. $\mathbf{e}$ is a vector selecting the first column of the speech covariance matrix and $E[\cdot]$ denotes mathematical expectation. 

Solving Eq. (\ref{eq:sc_mvdr}), the MF-MVDR filter vector can be obtained as \cite{huang2011multi,benesty2011single,tammen2019dnn}
\begin{equation}
\label{eq:sc_mvdr_solution_conven}
\mathbf{h_{\text{MF-MVDR}}}(t,f)=\frac{{\mathbf{\Phi}^{\text{MF}}_{\mathbf{VV}}}^{-1}(t,f) \mathbf{\boldsymbol{\gamma}_{x}}(t,f)}{\mathbf{\boldsymbol{\gamma}_{x}^{\mathrm{H}}}(t,f) {\mathbf{\Phi}^{\text{MF}}_{\mathbf{VV}}}^{-1}(t,f) \mathbf{\boldsymbol{\gamma}_{x}}(t,f)}. 
\end{equation}
Note that in \cite{tammen2019dnn}, $\mathbf{\Phi}^{\text{MF}}_{\mathbf{VV}}$ was replaced by the MF noise covariance matrix $\mathbf{\Phi}^{\text{MF}}_{\mathbf{NN}}$ under the assumption that the uncorrelated speech component is negligible, which imposes an upper bound on the system's performance.

\section{Proposed ADL-MVDR Beamformer}
\label{sec:Proposed ADL-MVDR Beamformer}

As mentioned in previous sections, the covariance matrices for most of the conventional neural mask-based MC-MVDR systems are computed at the chunk-level that discards the temporal information. This results in relatively high levels of residual noise in the separated speech. Additionally, the matrix inverse involved in conventional neural mask-based MVDR systems is sometimes not numerically stable when jointly trained with NNs~\cite{zhang2021end}. In some classical MVDR approaches, the derivations of the steering or IFC vectors and covariance matrices are based on recursive methods which requires heuristic updating factors between consecutive frames~\cite{souden2011integrated,schwartz2014multi,tammen2019dnn}. However, these factors are usually hard to determine and could easily influence the accuracy of the estimated terms. 

In this study, we propose a novel idea of ADL-MVDR that can be applied to many different configurations, including MC beamforming, MF filtering and MCMF beamforming (note that a separate system needs to be trained for each microphone setup). The key idea of our proposed ADL-MVDR framework is about using two separate gated recurrent unit (GRU) based networks (denoted as GRU-Nets) to replace the matrix inverse and principal eigenvector extraction processes involved in MVDR solution. Leveraging on the temporal properties of RNNs, the GRU-Nets can better explore and utilize the temporal information from previous frames without any needs of heuristic updating factors. Estimating MVDR coefficients via GRU-Nets also bypasses the numerical instability issue caused by the matrix inverse, whereas conventional methods often use diagonal loading and gradient clipping to address this issue~\cite{xu2020neural, zhang2021end}. Note that previous approaches that adopt NNs to directly learn the beamforming filtering weights \cite{xiao2016study,li2016neural} are not successful since noise information is not explicitly considered, whereas our proposed ADL-MVDR beamformer explicitly utilizes the cross-channel information from both estimated speech and noise covariance matrices. 

\begin{figure}[t!]
  \centering
  \includegraphics[scale = 0.35]{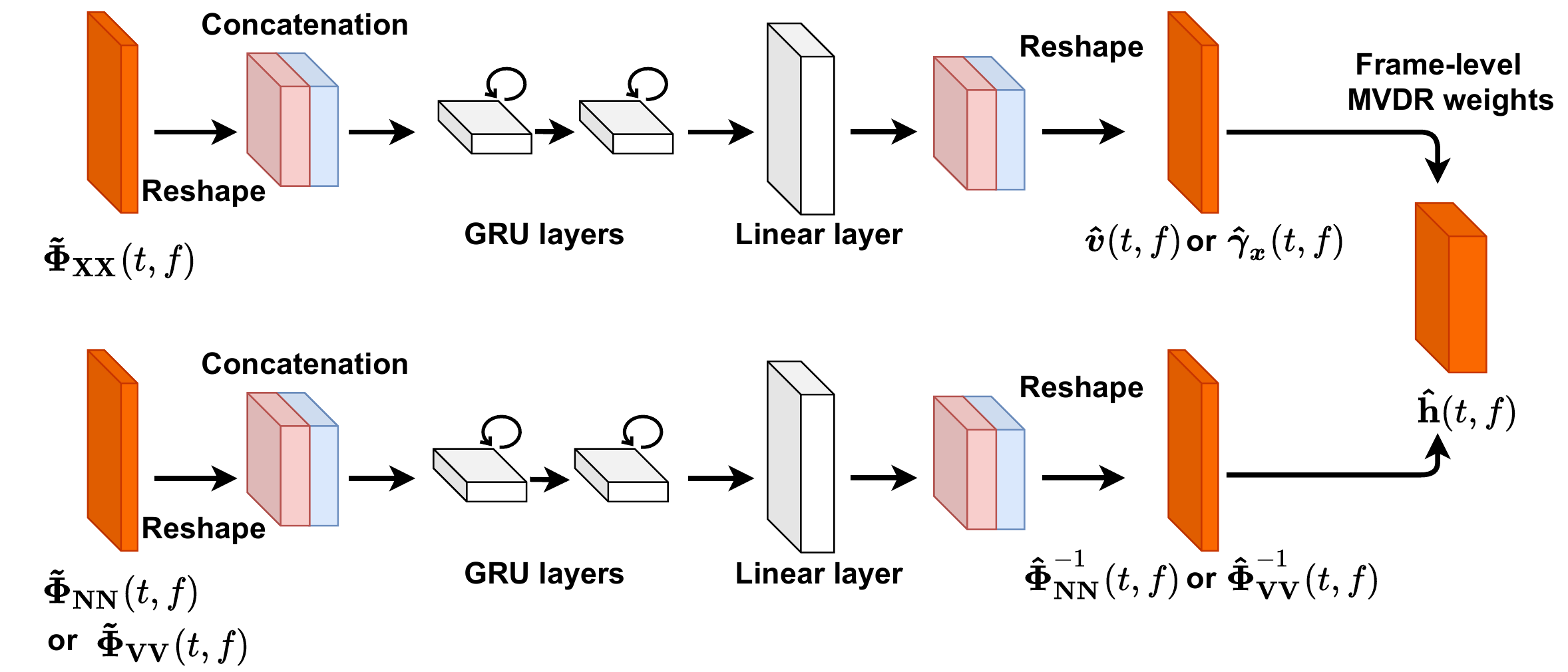}
  \caption{(Color Online). Detailed network architecture of the ADL-MVDR module. The real and imaginary parts of the covariance matrices are concatenated before fed into the GRU-Nets. The estimated MVDR coefficients are reshaped back to their original forms before computing the frame-level MVDR weights.}
  \label{fig:ADL-MVDR module}
\end{figure}

\subsection{System Overview}
\label{subsec:System Overview}

The general framework of our proposed ADL-MVDR beamformer is depicted in Fig. \ref{fig:ADL-MVDR}. The system consists of two parts, a complex filter estimator that is based on our previously proposed multi-modal MC speech separation platform \cite{tan2020audio,gu2020multi} (i.e., a Conv-TasNet \cite{luo2019conv} variant) for covariance matrices estimation, followed by another ADL-MVDR module (depicted in Fig. \ref{fig:ADL-MVDR module}) for frame-level MVDR weights derivation. 

\begin{figure}[t!]
  \includegraphics[scale = 0.26]{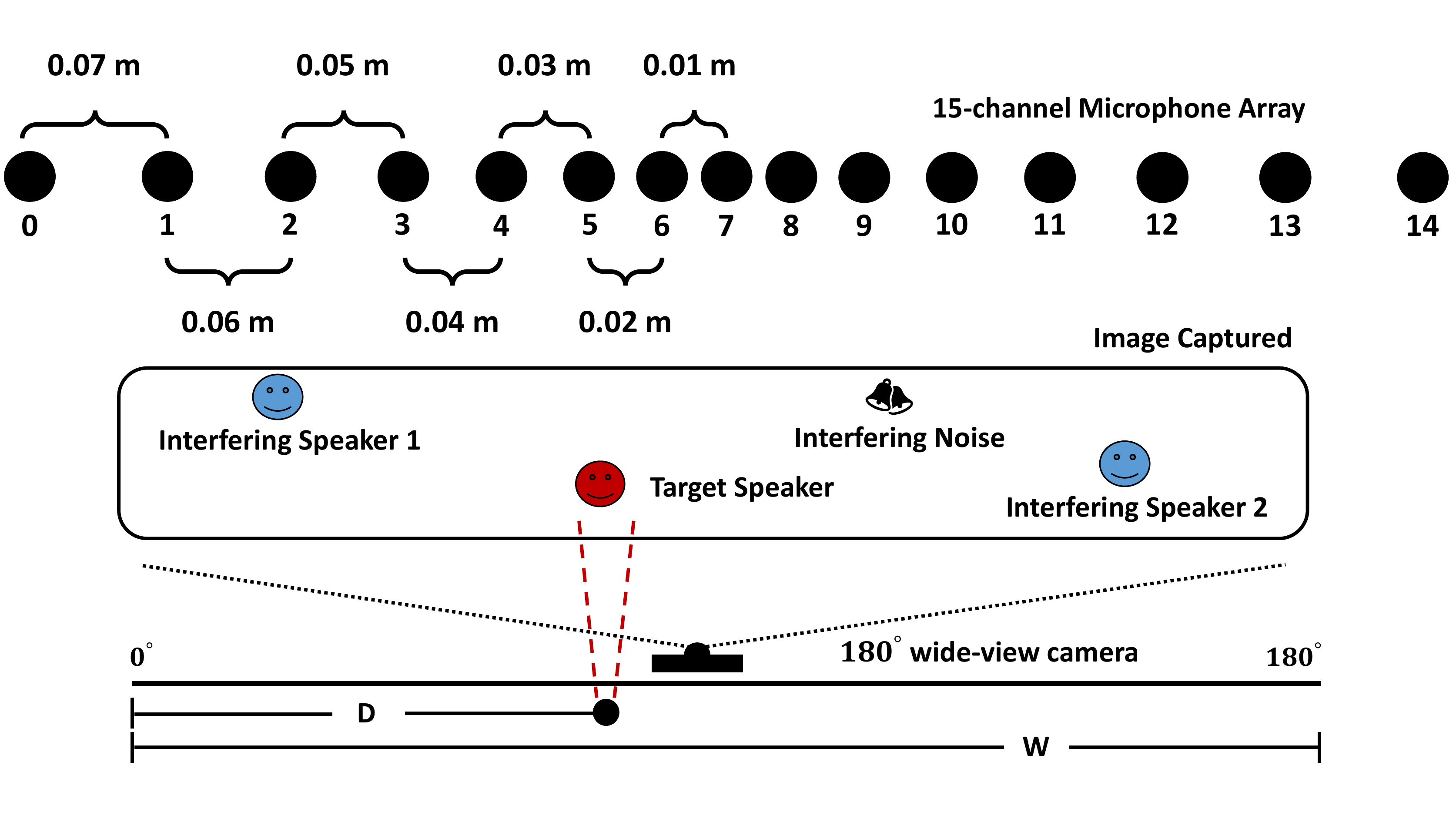}
  \centering
  \caption{(Color Online). Schematic diagrams for the 15-channel linear microphone array and the DOA estimation process with a wide view camera.}
  \label{fig:mic_array}
\end{figure}

As described in our previous works \cite{gu2020multi,tan2020audio}, inside the complex filter estimator, the interaural phase difference (IPD) and log-power spectra (LPS) features are extracted from the 15-channel noisy speech, where the IPD between the $p$-th pair of microphone channels (e.g., $m_1$ and $m_2$) is
\begin{equation}
\begin{aligned}
\label{eq:IPD_calculation}
\text{IPD}^{(p)}(t,f) = \angle \mathrm{Y}^{(m_1)}(t,f) - \angle \mathrm{Y}^{(m_2)}(t,f),
\end{aligned}
\end{equation}
where $\angle$ extracts the phase angle. We use the IPD computed from five pairs of channels, i.e., (0,14), (1,13), (2,11), (4,11), and (6,8), corresponding to five different distances between microphones \cite{tan2020audio,gu2020multi}. A directional feature (DF) \cite{chen2018multi} is also used in our experiment, which is defined as the cosine distance between the steering vector and IPD:
\begin{equation}
\begin{aligned}
\label{eq:DF_calculation}
\text{DF}(t,f) &= \sum_{p=1}^{P} \langle \mathbf{e}^{\text{TPD}^{(p)}(\theta_{t},f)}, \mathbf{e}^{\text{IPD}^{(p)}(t,f)}\rangle, \\
\text{TPD}^{(p)}(\theta_{t},f) &= 2\pi f \Delta_{p} \cos \theta_{t} / (f_s c),
\end{aligned}
\end{equation}
where $\langle \rangle$ denotes cosine distance and vector $\mathbf{e}^{(\cdot)}=\left[\begin{array}{l}\cos (\cdot) \\ \sin (\cdot)\end{array}\right]$. $\text{TPD}^{(p)}(\theta_{t},f)$ is the target-dependent phase difference that describes the phase delay of a plane wave (with frequency $f$), at the $p$-th pair of microphones (total number of $P$ pairs), with target direction of $\theta_{t}$, $\Delta_{p}$ is the distance between the $p$-th pair of microphones, $f_s$ is the sampling frequency and $c$ is the sound velocity. The target DOA (i.e., $\theta_{t})$ can be estimated using a wide-view camera by locating the target speaker's face. Once the DF is obtained, it is further merged with the IPD and LPS features by concatenating along the feature dimension to form the input vector (1799 feature size) to the separation network \cite{tan2020audio,gu2020multi}. An illustration of the 15-channel microphone array that is calibrated with a $180^{\circ}$ camera is depicted in Fig. \ref{fig:mic_array}, where the microphone array is a linear array and is symmetric to the center microphone (i.e., the 7th microphone in Fig. \ref{fig:mic_array}). The DOA is roughly estimated with the target source location D on the image with width W that is captured by the camera, where $\text{DOA}= \mathrm{\frac{D}{W}}\times180^{\circ}$.

In the front-end complex filter estimator, the audio encoding network first reduces the feature dimension of the input vector using a 1-D convolution layer with 256 1$\times$1 kernels (i.e., pointwise convolution) \cite{tan2020audio}, followed by a stack of two successive TCN blocks that consists of dilated 1-D convolution layers with dilation rates exponentially increased from $2^0$ to $2^7$. The details of TCN are illustrated in Fig. \ref{fig:audioencodingblock}. Within each dilated 1-D convolution block, the number of output channels is set to 512 for the first 1$\times$1 convolution layer, the depth-wise separable convolution layer has a kernel size of 3 with 512 output channels, followed by another 1$\times$1 convolution layer with 256 output channels. After audio encoding network, two separate filter estimator networks are used to estimate the speech and noise (or undesired signal) components. Each filter estimator network consists of two successive TCN blocks (with dilation rates from $2^0$ to $2^7$) followed by another 1$\times$1 convolution layer, where the number of output channels for the last 1$\times$1 convolution layer depends on the estimation target (i.e., cRM or cRF). 

\subsection{Complex ratio filtering}
\label{subsec:complex ratio filtering}

Our recent work \cite{xu2020neural} suggests that the complex ratio mask (denoted as cRM) can lead to better system performance, the procedure of using cRM to derive the covariance matrix is described below as
\begin{equation}
\begin{aligned}
\label{eq:crm_cov_mat}
\mathbf{\hat{X}_{cRM}}(t,f) &=  (\mathrm{cRM_{r}} + j\mathrm{cRM_{i}})\cdot(\mathrm{\mathbf{Y}_{r}} + j\mathrm{\mathbf{Y}_{i}}) \\
&= \mathrm{cRM}_{\mathrm{X}}(t,f) \cdot \mathbf{Y}(t,f), \\
\mathbf{\hat{\Phi}}_{\mathbf{XX}}(f) &= \frac{\sum_{t=1}^{T} \mathbf{\hat{X}_{cRM}}(t,f) \mathbf{\hat{X}_{cRM}}^{\mathrm{H}}(t,f)}{\sum_{t=1}^{T} \mathrm{cRM}_{\mathrm{X}}^{\mathrm{H}}(t,f) \mathrm{cRM}_{\mathrm{X}}(t,f)} ,
\end{aligned}
\end{equation}
where $\mathbf{\hat{X}_{cRM}}(t,f)$ represents the estimated MC speech component via the complex speech mask $\mathrm{cRM}_{\mathrm{X}}$. $r$ and $i$ denote the real and imaginary parts, respectively. `$\cdot$' is the complex multiplier and $j$ is the complex number. The power of the complex mask is used for normalization.

\begin{figure}[t!]
  \centering
  \includegraphics[scale = 0.5]{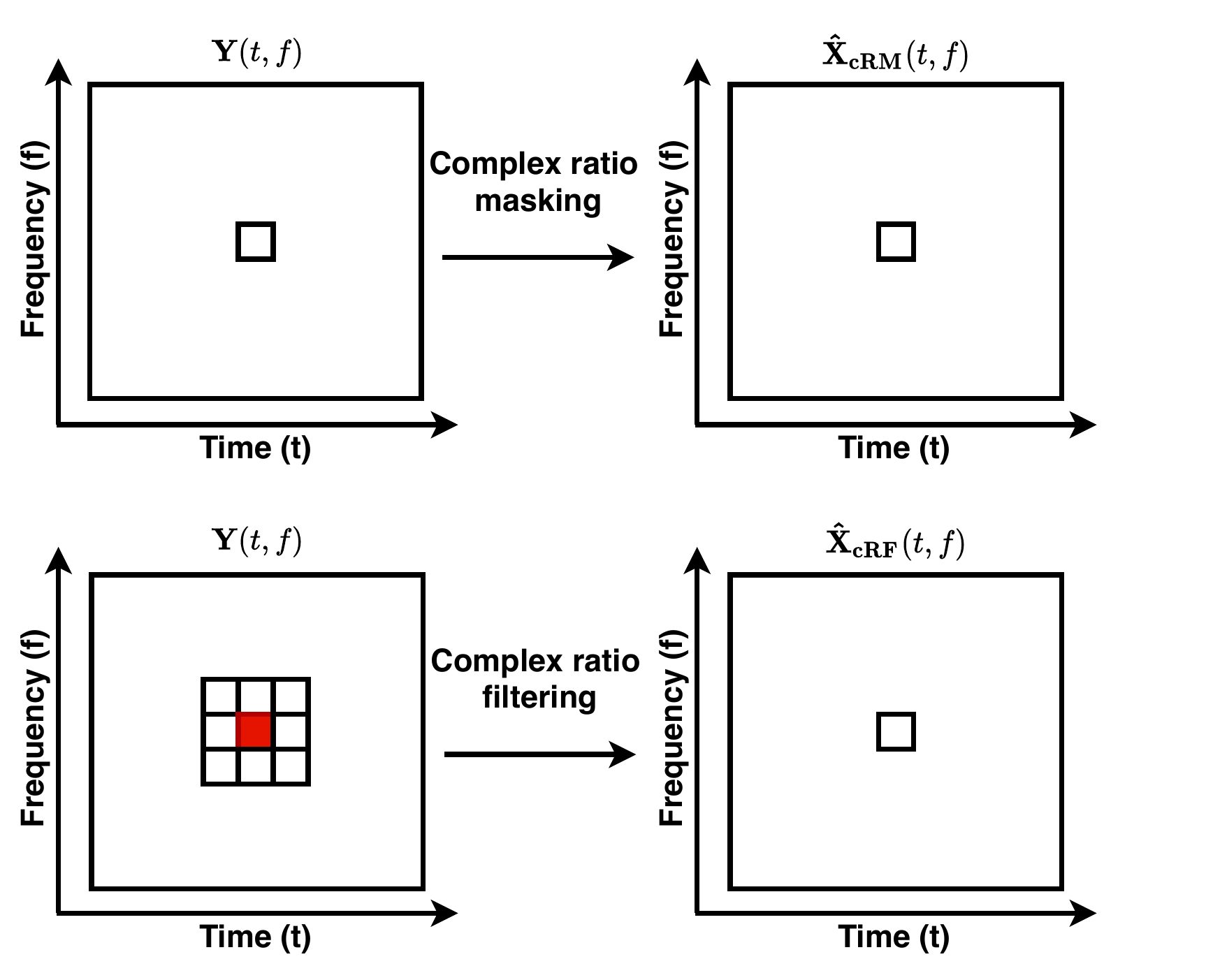}
  \caption{(Color Online). Schematic diagrams for complex ratio masking and complex ratio filtering. The cRM is a one-to-one mapping, whereas the cRF is a many-to-one mapping. In this example, the cRF has a size of 3$\times$3, where each T-F bin is estimated using nine T-F bins in its neighboring filter range. The center mask is marked with a red fill.}
  \label{fig:cRF_plot}
\end{figure}

In this study, different from prior neural mask-based MVDR approaches that use T-F masks (e.g., ideal ratio mask (IRM), cRM, etc.) to estimate the speech and noise/undesired signal components, we adopt the cRF \cite{mack2019deep} method for estimation. As depicted in Fig. \ref{fig:cRF_plot}, the cRF differs from the cRM that instead of using one-to-one mapping, it utilizes the nearby T-F bins to estimate each target T-F bin. The example shown in Fig. \ref{fig:cRF_plot} can be formulated as
\begin{equation}
\begin{aligned}
\label{eq:cRF_filtering}
\mathbf{\hat{X}}_\mathrm{\mathbf{cRF}}(t,f) &= \sum_{\tau_{1}=-J_{1}}^{J_{2}}\sum_{\tau_{2}=-K_{1}}^{K_{2}}\mathrm{cRF}(t,f,\tau_{1},\tau_{2})\\
&\cdot\mathbf{Y}(t+\tau_{1},f+\tau_{2}), \\
\end{aligned}
\end{equation}
where $\mathbf{\hat{X}}_\mathrm{\mathbf{cRF}}$ is the estimated MC speech component using the cRF method, the cRF has a size of $(J_{2}+J_{1}+1) \times (K_{2}+K_{1}+1)$ for each T-F pixel in the estimated speech. $J_{1}, J_{2}$ and $K_{1}, K_{2}$ represent the number of previous/lower and future/higher frames and frequency bins used for filtering, respectively. The noise or the undesired signal components can be obtained in a similar manner. Different from \cite{mack2019deep}, where the authors directly apply cRF for speech separation, we adopt cRF to estimate the speech and noise covariance matrices which are further used as inputs to the ADL-MVDR module. The cRF also plays an important role in the success of our ADL-MVDR beamformer, which will be illustrated afterwards in the ablation study.

\subsection{Multi-channel ADL-MVDR}
\label{subsec:Multi-channel ADL-MVDR}

An accurate estimate of the steering vector is very important for an MC-MVDR system as it contains information about which direction the signal should be preserved \cite{Griffiths1982AnAA,doclo2002gsvd}. Yet previous approaches involving extracting the principal eigenvector on the speech covariance matrix sometimes could introduce large gradients during back propagation and gradient clipping is often adopted \cite{heymann2015blstm,xu2019joint,xu2020neural}. A similar issue exists for the inversion process of the estimated noise covariance matrix when it becomes singular during joint training \cite{zhao2012fast,lim2017numerical, zhang2021end}, where diagonal loading~\cite{mestre2003diagonal} is often used to stabilize the process \cite{chen2013finite,xu2019joint,xu2020neural,zhang2021end}. In order to perform time-varying beamforming and stabilize the joint training process, we deploy two GRU-Nets to replace the principal eigenvector extraction and matrix inverse involved in MVDR solution. Note that we concatenate the real and imaginary parts of the NN estimated time-varying covariance matrices (i.e., $\mathbf{\tilde{\Phi}}_{\mathbf{XX}}$ and $\mathbf{\tilde{\Phi}}_{\mathbf{NN}}$) before feeding them into the GRU-Nets, as shown in Fig. \ref{fig:ADL-MVDR module}. 
\begin{equation}
\begin{aligned}
\label{eq:GRU_coef}
\mathbf{\hat{\boldsymbol{v}}}(t,f) & = \mathbf{GRU{\text -}Net}_{\boldsymbol{v}}(\mathbf{\tilde{\Phi}}_{\mathbf{XX}}(t^{\prime},f)), \\
\mathbf{\hat{\Phi}}_{\mathbf{NN}}^{-1}(t,f) & = \mathbf{GRU{\text -}Net}_{\boldsymbol{\mathrm{NN}}}(\mathbf{\tilde{\Phi}}_{\mathbf{NN}}(t^{\prime},f)), \\
\end{aligned}
\end{equation}
where $0 \leq t^{\prime} \leq t$. Leveraging on the temporal properties of RNNs, the frame-wise covariance matrices are fed into the GRU-Nets for MVDR coefficients estimation. The hidden internal states of the GRU-Nets can help capture the previous temporal information in the estimated MVDR coefficients (i.e., steering vector and inverse of the noise covariance matrix). Without using any arbitrary inter-frame updating factors, the GRU-Nets can learn the temporal dependencies through the NN training process. The input time-varying speech and noise covariance matrices (note they are frame-wise matrices without taking expectation over time) can be obtained from the cRF estimated speech and noise components as
\begin{equation}
\begin{aligned}
\label{eq:eq:multi-channel-framewise-cov-mat}
\mathbf{\tilde{\Phi}}_{\mathbf{XX}}(t,f) &=\frac{\mathbf{\hat{X}}_\mathrm{\mathbf{cRF}}(t,f) \mathbf{\hat{X}}_\mathrm{\mathbf{cRF}}^{\mathrm{H}}(t,f)}{\sum_{t=1}^{T}\mathrm{cRM}_{\mathrm{X}}^{\mathrm{H}}(t,f) \mathrm{cRM}_{\mathrm{X}}(t,f)},\\
\mathbf{\tilde{\Phi}}_{\mathbf{NN}}(t,f) &=\frac{\mathbf{\hat{N}}_\mathrm{\mathbf{cRF}}(t,f) \mathbf{\hat{N}}_\mathrm{\mathbf{cRF}}^{\mathrm{H}}(t,f)}{\sum_{t=1}^{T}\mathrm{cRM}_{\mathrm{N}}^{\mathrm{H}}(t,f) \mathrm{cRM}_{\mathrm{N}}(t,f)},
\end{aligned}
\end{equation}
where $\mathbf{\hat{N}}_\mathrm{\mathbf{cRF}}$ and $\mathrm{cRM}_{\mathrm{N}}$ denote the estimated MC noise component using the cRF method and the center complex mask of the noise cRF (as depicted in Fig. \ref{fig:cRF_plot}) that is used for normalization, respectively. The same notation also applies to $\mathrm{cRM}_{\mathrm{X}}$. Different from Eq. (\ref{eq:crm_cov_mat}), we do not sum over the temporal dimension to preserve the frame-level information as input to the GRU-Nets. 

Based on these RNN-derived frame-wise MVDR coefficients, i.e., $\mathbf{\hat{\Phi}}^{-1}_{\mathbf{NN}}(t,f) \in \mathbb{C}^{M \times M}$ and $\mathbf{\hat{\boldsymbol{v}}}(t,f) \in \mathbb{C}^{M}$, the MC MVDR beamforming weights can be derived at frame-level as
\begin{equation}
\begin{aligned}
\label{MVDR_tf_solutions}
\mathbf{\hat{h}_{\text{MC ADL-MVDR}}}(t,f) & = \frac{\mathbf{\hat{\Phi}}_{\mathbf{NN}}^{-1}(t,f) \hat{\boldsymbol{v}}(t,f)}{\mathbf{\hat{\boldsymbol{v}}^{\mathrm{H}}}(t,f) \mathbf{\hat{\Phi}}_{\mathbf{N} \mathbf{N}}^{-1}(t,f) \mathbf{\hat{\boldsymbol{v}}}(t,f)},
\end{aligned}
\end{equation}
where $\mathbf{\hat{h}_{\text{MC ADL-MVDR}}}(t,f) \in \mathbb{C}^{M}$ is the frame-wise MC ADL-MVDR beamforming weights which are different from the chunk-level weights derived in many conventional neural mask-based MC-MVDR systems. Finally, the MC ADL-MVDR enhanced speech is obtained as
\begin{equation}
\label{eq:tf-beamforming}
\mathrm{\hat{X}^{(0)}_{\text{MC ADL-MVDR}}}(t,f) = \mathbf{\hat{h}^{\mathrm{H}}_{\text{MC ADL-MVDR}}}(t,f)\mathbf{Y}(t,f).
\end{equation}

\subsection{Multi-frame ADL-MVDR}
\label{subsec:Multi-frame ADL-MVDR}

We also introduce an MF setup to our ADL-MVDR framework to simulate an extreme condition when only one channel of the signal is available to the beamforming module. Note that the MF ADL-MVDR system still uses the MC noisy speech as inputs to estimate the cRFs, however, we only use one channel of the signal as the inputs to the ADL-MVDR module. Without a loss of generality, the front-end complex filter estimator could be replaced by any other speech separation systems. The network architecture of the MF ADL-MVDR system is analogous to the MC ADL-MVDR, however, the purpose for each step in this case is very different. Since the spatial information from the microphone array is no longer available, the MF-MVDR explores the correlation information between consecutive frames instead. An accurate estimate on the speech IFC vector dominates the final performance of the system. 

Similar to the MC case, the time-varying MF speech covariance matrix can be derived based on the estimated MF speech component. They are defined as
\begin{equation}
\begin{aligned}
\label{eq:multi-frame_cov_mat}
\mathbf{\hat{\overline{X}}^{(0)}}_\mathrm{\mathbf{cRF}}(t,f) &= [\mathrm{\hat{X}^{(0)}_{cRF}}(t-L_1+1,f),...,\mathrm{\hat{X}^{(0)}_{cRF}}(t,f),\\
&...,\mathrm{\hat{X}^{(0)}_{cRF}}(t+L_2,f)]^{T},\\
\mathbf{\tilde{\Phi}}^{\text{MF}}_{\mathbf{XX}}(t,f) &=\frac{\mathbf{\hat{\overline{X}}^{(0)}}_\mathrm{\mathbf{cRF}}(t,f) \mathbf{\hat{\overline{X}}^{(0)\: \mathrm{H}}}_\mathrm{\mathbf{cRF}}(t,f)}{\sum_{t=1}^{T}\mathrm{cRM}_{\mathrm{X}}^{\mathrm{H}}(t,f) \mathrm{cRM}_{\mathrm{X}}(t,f)},
\end{aligned}
\end{equation}
where $\mathbf{\hat{\overline{X}}^{(0)}}_\mathrm{\mathbf{cRF}}$ is the estimated $L$-frame ($L = L_1+L_2$, where $L_1$ and $L_2$ indicate the number of previous and future frames) single-channel speech component in T-F domain using the cRF method. $\mathrm{\hat{X}^{(0)}_{cRF}}$ is the estimated single-channel speech component via cRF. The time-varying covariance matrix of the undesired signal $\mathbf{\tilde{\Phi}}^{\text{MF}}_{\mathbf{VV}}$ can be estimated in a similar way using cRF method. Note that the undesired signal component is estimated implicitly as the neural network can gradually learn the mapping during training. 

In the MF ADL-MVDR system, two GRU-Nets are implemented to estimate the speech IFC vector and the inverse of the undesired signal covariance matrix. The inputs to these two networks are the time-varying MF speech and undesired signal covariance matrices $\mathbf{\tilde{\Phi}}^{\text{MF}}(t,f) \in \mathbb{C}^{L \times L}$ estimated via the cRF method. This is formulated as
\begin{equation}
\begin{aligned}
\mathbf{\boldsymbol{\hat{\gamma}}_{x}}(t,f) & = \mathbf{GRU{\text -}Net}_{\boldsymbol{\gamma}}(\mathbf{\tilde{\Phi}}^{\text{MF}}_{\mathbf{XX}}(t^{\prime},f)), \\
{\mathbf{\hat{\Phi}}^{\text{MF}\: -1}_{\mathbf{VV}}}(t,f) & = \mathbf{GRU{\text -}Net}_{\boldsymbol{\mathrm{VV}}}(\mathbf{\tilde{\Phi}}^{\text{MF}}_{\mathbf{VV}}(t^{\prime},f)).
\end{aligned}
\end{equation}
Once these variables are obtained, the MF ADL-MVDR filter weights $\mathbf{\hat{h}_{\text{MF ADL-MVDR}}}(t,f) \in \mathbb{C}^{L}$ are applied to the MF noisy speech $\mathbf{\overline{Y}^{(0)}}(t,f) \in \mathbb{C}^{L}$ and the estimated speech are obtained as
\begin{equation}
\label{eq:sc_mvdr_solution}
\begin{aligned}
\mathbf{\hat{h}_{\text{MF ADL-MVDR}}}(t,f) &=\frac{\mathbf{\hat{\Phi}}^{\text{MF}\: -1}_{\mathbf{VV}}(t,f) \mathbf{\boldsymbol{\hat{\gamma}}_{x}}(t,f)}{{\boldsymbol{\hat{\gamma}}^{\mathrm{H}}_{x}}(t,f) \mathbf{\hat{\Phi}}^{\text{MF}\: -1}_{\mathbf{VV}}(t,f) \mathbf{\boldsymbol{\hat{\gamma}}_{x}}(t,f)}, \\
\mathrm{\hat{X}^{(0)}_{\text{MF ADL-MVDR}}}(t,f) &= \mathbf{\hat{h}^{\mathrm{H}}_{\text{MF ADL-MVDR}}}(t,f)\mathbf{\overline{Y}^{(0)}}(t,f).
\end{aligned}
\end{equation}

\subsection{Multi-channel Multi-frame ADL-MVDR}
\label{subsec:Multi-channel Multi-frame ADL-MVDR}

The MCMF ADL-MVDR combines both the MC and MF information and uses them as inputs to the ADL-MVDR module. Let $\mathbf{\hat{X}_{cRF}}$ denote the MC speech estimated using the cRF method, then the MCMF speech and its time-varying covariance matrix are obtained as
\begin{equation}
\begin{aligned}
\label{eq:multi-channal-multi-frame_cov_mat}
\mathbf{\hat{\overline{X}}}_\mathrm{\mathbf{cRF}}(t,f) &= [\mathbf{\hat{X}_{cRF}}(t-L_1+1,f),...,\mathbf{\hat{X}_{cRF}}(t,f),\\
&...,\mathbf{\hat{X}_{cRF}}(t+L_2,f)]^{T}, \\
\mathbf{\tilde{\Phi}}^{\text{MCMF}}_{\mathbf{XX}}(t,f) &=\frac{\mathbf{\hat{\overline{X}}}_\mathrm{\mathbf{cRF}}(t,f) \mathbf{\hat{\overline{X}}}_\mathrm{\mathbf{cRF}}^{\mathrm{H}}(t,f)}{\sum_{t=1}^{T}\mathrm{cRM}_{\mathrm{X}}^{\mathrm{H}}(t,f) \mathrm{cRM}_{\mathrm{X}}(t,f)}.
\end{aligned}
\end{equation}
The MCMF estimated noise $\mathbf{\hat{\overline{N}}_{cRF}}$ component and its time-varying covariance matrix $\mathbf{\tilde{\Phi}}^{\text{MCMF}}_{\mathbf{NN}}$ can be estimated using the same method. Once the time-varying speech and noise covariance matrices $\mathbf{\tilde{\Phi}}^{\text{MCMF}}(t,f) \in \mathbb{C}^{ML \times ML}$ are obtained, we can follow similar steps as the MC ADL-MVDR described in Eq. (\ref{eq:GRU_coef}) to estimate the MCMF steering vector $\mathbf{\hat{\boldsymbol{v}}}^{\text{MCMF}}(t,f) \in \mathbb{C}^{ML}$ and inverse of the MCMF noise covariance matrix. After that, the MCMF ADL-MVDR beamforming weights $\mathbf{\hat{h}_{\text{MCMF ADL-MVDR}}}(t,f) \in \mathbb{C}^{ML}$ are derived in a similar manner as described in Eq. (\ref{MVDR_tf_solutions}). Finally, the MCMF ADL-MVDR enhanced speech $\mathrm{\hat{X}^{(0)}_{\text{MCMF ADL-MVDR}}}$ is obtained
\begin{equation}
\label{eq:tf-mcmfbeamforming}
\mathrm{\hat{X}^{(0)}_{\text{MCMF ADL-MVDR}}}(t,f) = \mathbf{\hat{h}^{\mathrm{H}}_{\text{MCMF ADL-MVDR}}}(t,f)\mathbf{\overline{Y}}(t,f),
\end{equation}
where $\mathbf{\overline{Y}}(t,f) \in \mathbb{C}^{ML}$ is the MCMF noisy speech.

\section{Experimental Setup}
\label{sec:Experimental Setup}

\subsection{Speech Materials}
\label{subsec:Speech Materials}

We adopt a Mandarin audio-visual speech corpus (will be released soon) collected from Youtube, which has been reported in our prior works \cite{xu2020neural,tan2020audio,gu2020multi}. An SNR estimator together with a face detector is used to filter out the low-quality ones \cite{tan2020audio,gu2020multi}. A total number of 205,500 clean video segments from around 1500 speakers are gathered. In contrast to our prior works \cite{tan2020audio,gu2020multi}, we do not use the lip movement features for our system since we focus on beamforming in this study. There are 190,000 speech utterances in the training set, 15,000 utterances in the validation set, and another 455 utterances in the testing set. Speakers in the testing set are different from those in the training set. The sampling rate is set to 16 kHz and the noise source contains random clips from 255 noises recorded indoors. 

To generate the multi-channel data, the microphone array and all the sound sources are randomly placed in a simulated room, the distance between any sound source and the microphone array is between 0.5 m to 6 m \cite{tan2020audio}. The number of overlapped speakers ranges from 1 to 3. We also consider a wide range of reverberant conditions, where 2000 different rooms with 6000 room impulse responses (RIRs) are simulated via the image method \cite{allen1979image}. The room size is randomly chosen from 4 m $\times$ 4 m $\times$ 3 m to 10 m $\times$ 10 m $\times$ 6 m, with RT60s ranging from 0.05 s to 0.7 s. As a result, the simulated multi-channel speech data has an SNR range of 18 to 30 dB and the signal-to-interference ratio (SIR) is between -6 to 6 dB.

\subsection{Audio Features and Training Procedure}
\label{subsec:Audio Features}

In order to extract the audio features, we use a 512-point STFT together with a 32 ms Hann window and 16 ms step size. During the training stage, the batch size and audio chunk size are set to 12 and 4 s, respectively. We adopt Adam optimizer with the initial learning rate set to $1e^{-3}$, Pytorch 1.1.0 is used. All models are trained with 60 epochs and early stopping is applied. The entire system is trained to minimize the time-domain scale-invariant source-to-noise ratio (Si-SNR) loss \cite{le2019sdr}, which is the negative of Si-SNR, i.e.,
\begin{equation}
\begin{aligned}
\mathcal{L}_{\text{Si-SNR}} &= -20\text{log}_{10} \frac{\|\alpha \cdot \mathbf{x}\|}{\|\hat{\mathbf{x}}-\alpha \cdot \mathbf{x}\|},\\
\alpha &= \frac{\hat{\mathbf{x}}^{\text{T}}\mathbf{x}}{\mathbf{x}^{\text{T}}\mathbf{x}},
\end{aligned}
\end{equation}
where $\alpha$ is a scaling factor that ensures the scaling invariance, $\hat{\mathbf{x}}$ denotes the time-domain estimated speech. This time-domain loss fits our end-to-end training paradigm. Note that we use the reverberant clean speech as the learning target as we mainly focus on separation in this study, the systems are not trained for dereverberation in the present study.

\subsection{System Setups}
\label{subsec:System Setups}

We adopt a Conv-TasNet variant \cite{luo2019conv} for complex filter estimation, which contains bunch of dilated 1-D convolution networks together with a pair of fixed STFT/iSTFT encoder/decoder implemented with 1-D convolution layers~\cite{tan2020audio,gu2020multi}. Details on the audio encoding network is introduced in Section \ref{subsec:System Overview}. In this study, we focus on the ADL-MVDR framework which can be adapted for MC, MF and MCMF-MVDR filtering. Three setups corresponding to these three scenarios are described below.

For the MC ADL-MVDR system, both GRU-Nets consist of two layers of GRU and another fully connected layer. The $\mathbf{GRU{\text -}Net}_{\boldsymbol{\mathrm{NN}}}$ uses 500 units for both GRU layers and 450 units (i.e., $\text{\# of channel}\times\text{\# of channel}\times\text{real and imaginary parts}$ = 15$\times$15$\times$2) for the fully connected layer. The $\mathbf{GRU{\text -}Net}_{\boldsymbol{v}}$ contains 500 and 250 units for each GRU layer, respectively, followed by a fully connected layer with 30 units (i.e., 15$\times$2). Tanh activation function is used for all GRU layers and linear activation function is used for the fully connected layers. The cRF size is empirically set to 3$\times$3 (i.e., a T-F bin with its surrounding eight T-F bins as depicted in Fig. \ref{fig:cRF_plot}) where it can utilize temporal information from one previous frame to one future frame and nearby frequency information from one frequency bin below to one frequency bin above. Note that the number of the output channels of the final 1$\times$1 convolution layer in the filter estimator networks could be changed accordingly for different configurations of cRF sizes. For example, we set the number of output channels to 257$\times$9 for a 3 $\times$ 3 cRF.

In terms of the MF ADL-MVDR system, the GRU-Nets feature the same structure to the MC setup, but with different hidden sizes. We use an MF size of five, i.e., from two previous frames to two future frames. The size of the cRF is 3$\times$3, identical to the MC setup. Unit size of all GRU layers is set to 128. The fully connected layer contains 10 units for $\mathbf{GRU{\text -}Net}_{\boldsymbol{\gamma}}$ and 50 units for $\mathbf{GRU{\text -}Net}_{\boldsymbol{\mathrm{VV}}}$.  

To investigate the influence of incorporating additional MF information on top of the MC spatial information, a 9-channel (i.e., mics: 0, 2, 3, 5, 7, 9, 11, 12, 14) 3-frame (i.e., from one previous frame to one future frame) MCMF ADL-MVDR system is included. Here the $\mathbf{GRU{\text -}Net}_{\boldsymbol{\mathrm{NN}}}$ consists of two GRU layer with 500 units each, followed by another 1458-unit fully connected layer. The $\mathbf{GRU{\text -}Net}_{\boldsymbol{v}}$ contains two GRU layers with 500 and 250 units, respectively, with another fully connected layer of 54 units. The cRF size is also set to $3\times3$.

Meanwhile, we investigate several microphone and MF setups in the ablation study for MCMF ADL-MVDR systems, including when only three (i.e., mics: 0, 7, 14), seven (i.e., mics: 0, 3, 5, 7, 9, 11, 14), nine (identical to the one mentioned above) or all 15 channels are available to the ADL-MVDR module. The cRF sizes are all set to $3\times3$, the MF sizes are set to three (i.e., from one previous frame to one future frame) and two (i.e., one previous frame to current frame) for different MCMF ADL-MVDR systems as presented in Table \ref{tab:2}. We also include their corresponding MC ADL-MVDR systems (i.e., without additional MF information) with the same selected microphone channels for comparison approaches. 

In other ablation studies, we examine the influence of the cRF size on the performance of MC/MF ADL-MVDR systems. The effects of different MF sizes are also investigated for MF ADL-MVDR systems. The statistics on model size, speed and memory usage are further provided. These results are reported in the following section. Note that the current configurations of ADL-MVDR systems are not specifically designed for online processing, although the audio encoding network and the ADL-MVDR beamformer could be configured to a causal system for online processing.

\subsection{Evaluation Metrics}
\label{subsec:Evaluation Metrics}

A set of objective evaluation metrics are used to evaluate the systems' performance from different perspectives. These metrics include PESQ \cite{rix2001perceptual}, source-to-distortion ratio (SDR) \cite{vincent2006performance} for speech quality assessment and STOI for intelligibility estimation \cite{taal2011algorithm}. The Si-SNR score is also included as it has been utilized for many recent speech separation systems \cite{luo2019conv,bahmaninezhad2019comprehensive,xu2020neural}. Moreover, we use a Tencent commercial speech recognition API \cite{tencentASR} (based on deep feed-forward sequential memory networks \cite{zhang2018deep}) to measure the ASR accuracy. The transcript of the speech is manually labelled by human annotators.

\begin{table*}[t!]
\centering
\caption{Evaluation results for our proposed ADL-MVDR systems. The PESQ scores are presented in detailed conditions including angle between the closest interfering source and total number of speakers. The average scores of Si-SNR, SDR and STOI are given for brevity. The ASR accuracy is measured with the WER and the best scores are highlighted in \textbf{bold} fonts.}
\label{tab:1}
\scalebox{0.9}{
\begin{tabular}{l|cccc|ccc|c|c|c|c|c}
\hline \hline
\multicolumn{1}{c|}{Systems/Metrics} & \multicolumn{8}{c|}{PESQ $\in [-0.5,4.5]$} & Si-SNR (dB) & \multicolumn{1}{l|}{SDR (dB)} & \multicolumn{1}{l|}{STOI ($\%$)} & WER ($\%$) \\ \hline 
 & 0-15$^{\circ}$ & 15-45$^{\circ}$ & 45-90$^{\circ}$ & 90-180$^{\circ}$ & 1spk & 2spk & 3spk & Avg. & Avg. & Avg. & Avg. & Avg.  \\ \hline
Reverberant clean (reference) & 4.50 & 4.50 & 4.50 & 4.50 & 4.50 & 4.50 & 4.50 & 4.50 & $\infty$ & $\infty$ & 100  & 8.26 \\
Noisy Mixture & 1.88 & 1.88 & 1.98 & 2.03 & 3.55 & 2.02 & 1.77 & 2.16 & 3.39 & 3.50 & 70.8 & 55.14 \\ \hline
\multicolumn{13}{c}{Purely NNs and our proposed MF ADL-MVDR system} \\ \hline
NN with cRM & 2.72 & 2.92 & 3.09 & 3.07 & 3.96 & 3.02 & 2.74 & 3.07 & 12.23 & 12.73 & 88.6  & 22.49 \\ 
NN with cRF & 2.75 & 2.95 & 3.12 & 3.09 & 3.98 & 3.06 & 2.76 & 3.10 & 12.50 & 13.01 & 89.2 & 22.07 \\ \hline
\textbf{MF ADL-MVDR} & $\textbf{2.80}$ & $\textbf{2.99}$ & $\textbf{3.16}$ & $\textbf{3.11}$ & $\textbf{4.01}$ & $\textbf{3.10}$ & $\textbf{2.80}$ & $\textbf{3.14}$ & $\textbf{12.60}$ & $\textbf{13.17}$ & $\textbf{89.5}$  & $\textbf{19.57}$ \\ \hline
\multicolumn{13}{c}{Conventional neural mask-based MVDR systems and our proposed MC ADL-MVDR system} \\ \hline
MVDR with cRF - recursive \cite{higuchi2016robust} & 2.56 & 2.75 & 2.93 & 2.90 & 3.75 & 2.87 & 2.56 & 2.90 & 10.12 & 11.10 & 89.0 & 16.74 \\ 
MVDR with cRM \cite{xu2020neural} & 2.55 & 2.76 & 2.96 & 2.84 & 3.73 & 2.88 & 2.56 & 2.90 & 10.62 & 12.04 & 88.5  & 16.85 \\ 
MVDR with cRF & 2.55 & 2.77 & 2.96 & 2.89 & 3.82 & 2.90 & 2.55 & 2.92 & 11.31 & 12.58 & 88.9  & 15.91 \\ \hline
Multi-tap MVDR with cRM \cite{xu2020neural} & 2.70 & 2.96 & 3.18 & 3.09 & 3.80 & 3.07 & 2.74 & 3.08 & 12.56 & 14.11 & 91.5  & 13.67 \\ 
Multi-tap MVDR with cRF & 2.67 & 2.95 & 3.15 & 3.10 & 3.92 & 3.06 & 2.72 & 3.08 & 12.66 & 14.04 & 91.4  & 13.52 \\ \hline
\textbf{MC ADL-MVDR} & $\textbf{3.04}$ & $\textbf{3.30}$ & $\textbf{3.48}$ & $\textbf{3.48}$ & $\textbf{4.17}$ & $\textbf{3.41}$ & $\textbf{3.07}$ & $\textbf{3.42}$ & $\textbf{14.80}$ & $\textbf{15.45}$ & $\textbf{93.3}$  & \textbf{12.73} \\ \hline 
\multicolumn{13}{c}{Our proposed MCMF ADL-MVDR system} \\ \hline
\textbf{MCMF ADL-MVDR} & $\textbf{3.10}$ & $\textbf{3.35}$ & $\textbf{3.48}$ & $\textbf{3.51}$ & $\textbf{4.20}$ & $\textbf{3.47}$ & $\textbf{3.10}$ & $\textbf{3.46}$ & $\textbf{15.43}$ & $\textbf{16.03}$ & $\textbf{93.7}$  & $\textbf{12.31}$ \\ 
\hline \hline
\end{tabular}}
\end{table*}

\begin{figure*}[t]
  \centering
  \includegraphics[scale = 0.54]{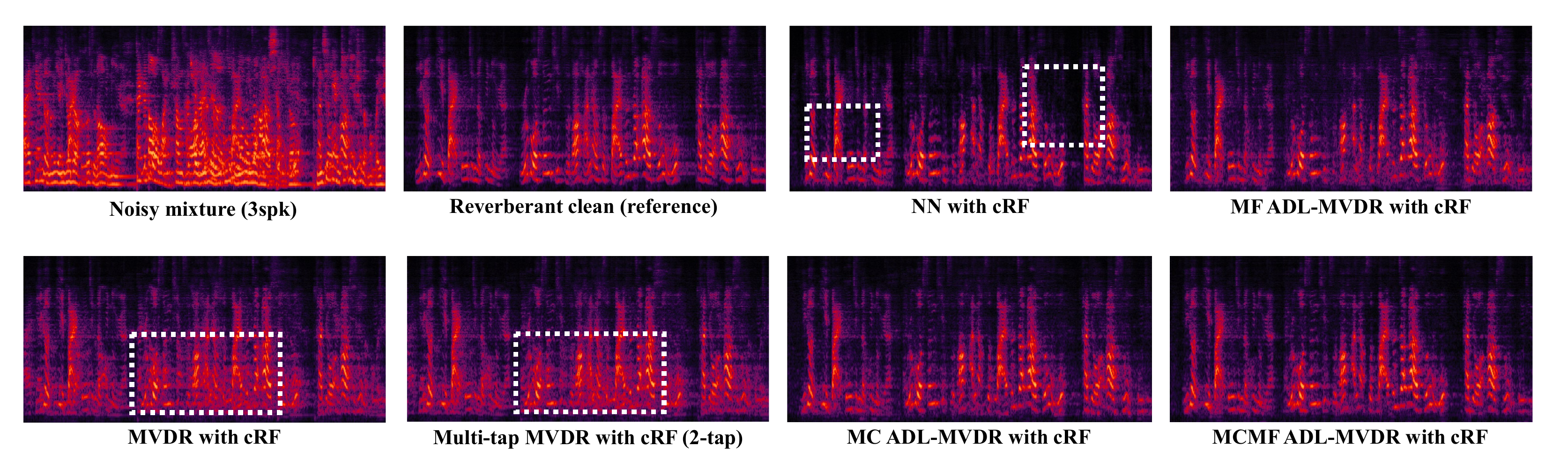}
  \caption{(Color Online). Spectrograms of some evaluated systems in Table \ref{tab:1}, the nonlinear distortion and residual noise are highlighted by the dashed boxes.}
  \label{fig:sample_spec}
\end{figure*}

\section{Results and Analysis}
\label{sec:Results and Analysis}

The general experimental results are provided in Table \ref{tab:1}, where we compare the performance of our proposed ADL-MVDR systems in MF, MC and MCMF conditions with its peers. Demos can be found on our website\footnote{Samples of separated speech (including real-world scenarios) are available at \href{https://zzhang68.github.io/mcmf-adl-mvdr/}{https://zzhang68.github.io/mcmf-adl-mvdr/}}. The performance of purely NN systems (i.e., the audio encoding network described in Section \ref{subsec:System Overview}, denoted as NN with cRM/cRF) are also included as baselines. We further include conventional neural mask-based MC-MVDR systems \cite{xu2019joint} and multi-tap MVDR systems \cite{xu2020neural} for comparison approaches (denoted as MVDR with cRM/cRF and multi-tap MVDR with cRM/cRF), they incorporate the same front-end network structure for mask/filter estimation as the ADL-MVDR systems, followed by conventional segment-level MVDR beamforming \cite{xu2019joint,xu2020neural}. Furthermore, we implement another MVDR systems with recursive updating rules for mini-block covariance matrices as described in \cite{higuchi2016robust,higuchi2017online} (denoted as MVDR with cRF - recursive), where the mini-block size is set to 30 frames with hop size of 10 frames. For fair comparison, the same front-end complex filter estimator was used for speech and noise covariance matrices estimation (with 3$\times$3 cRF), the center mask is used to compute the transition factors between successive mini-blocks. Note that all of our implemented conventional neural mask-based MVDR systems feature an end-to-end training scheme similar to \cite{heymann2017beamnet,subramanian2019investigation,xu2019joint,xu2020neural}, we further adopt diagonal loading~\cite{mestre2003diagonal} and gradient clipping \cite{heymann2015blstm,xu2020neural} to alleviate the numerical instability issue during joint training.

In Table \ref{tab:1}, the PESQ scores are further split up into detailed conditions, including the angle between the closest interfering source and total number of speakers. We present average results of other metrics (i.e., Si-SNR, SDR and STOI) for brevity. 

In the ablation studies, we report the performance with a set of speech evaluation metrics (e.g., PESQ, Si-SNR) as well as WER. The simulation results for different MCMF ADL-MVDR systems are provided in Table \ref{tab:2}. Table \ref{tab:3} illustrates the results on the effects of different cRF sizes on both MF and MC ADL-MVDR systems, where the filtering region for each T-F pixel is described by its relative boundaries of frames and frequency bins. The effects of MF sizes on the performance of MF ADL-MVDR systems are also revealed in Table \ref{tab:4}. We want to point out that there are infinitely many combinations of different cRF sizes and MF sizes, we only investigate a limited number of them which we consider to be representative. Lastly, Table \ref{tab:5} provides statistics on the model size, running speed and memory usage for a set of evaluated systems.

\subsection{Overview Results on ADL-MVDR Systems}
\label{subsec:Multi-channel and Multi-frame ADL-MVDR Systems}

\textbf{MC ADL-MVDR vs. Conventional neural mask-based MVDR}: we first investigate the performance of our proposed ADL-MVDR framework in the MC scenario. As provided in the third block of Table \ref{tab:1}, the MC ADL-MVDR system outperforms conventional neural mask-based MVDR systems by a large margin across all objective scores. For instance, in terms of the speech quality, our proposed MC ADL-MVDR system outperforms the MVDR system with cRF for more than 17\% (i.e., PESQ: 3.42 vs. 2.92). Even under extreme conditions when the interfering sources are very close to the target speaker (i.e., angles less than $15^{\circ}$), the MC ADL-MVDR system can still restore the separated speech quality to a high level (i.e., PESQ: 3.04). In terms of the Si-SNR and SDR performance, our proposed MC ADL-MVDR system also achieves nearly 31\% and 23\% improvements over the baseline MVDR system with cRF (i.e., Si-SNR: 14.80 dB vs. 11.31 dB, SDR: 15.45 dB vs. 12.58 dB). Similar patterns could be found for intelligibility scores, where our proposed MC ADL-MVDR system achieves around 5\% better performance than recursive and chunk-level MVDR systems (e.g., STOI: 93.3\% vs. 89.0\% and 88.9\%), respectively.

Compared to MVDR with cRF (i.e., chunk-level covariance estimation), the MVDR with recursive updating rules results in slightly worse performance (e.g., PESQ; 2.90 vs. 2.92, Si-SNR: 10.12 dB vs. 11.31 dB), which might be caused by the less accurate estimations of the covariance matrices within a mini-block rather than using a full chunk of audio. Yet it still achieves significant ASR gains compared to the purely NNs (i.e., 16.74\% vs. 22.49\% and 22.07\%). Note that the residual noise in recursive MVDR and conventional neural mask-based MVDR systems is still at relatively high level according to their PESQ scores (i.e., average scores of 2.90 and 2.92, respectively) when compared to the purely NN systems (e.g., NN with cRF: 3.10). The multi-tap MVDR with cRF can better remove the residual noise than MVDR with cRF, however, the residual noise is still high compared to MC ADL-MVDR system. Our proposed MC ADL-MVDR system also demonstrates its superiority in ASR accuracy when compared to the multi-tap MVDR system with cRF (i.e., 12.73\% vs. 13.52\%). Considering that the current commercial ASR system is already very robust to low-level of noises, the nearly 6\% relative improvement in WER is fairly good since our MC ADL-MVDR system can greatly remove the residual noise while introducing even fewer distortions to the target speech simultaneously. This can also be observed in the example spectrograms provided in Fig. \ref{fig:sample_spec}, the conventional neural mask-based MVDR and multi-tap MVDR systems come with high levels of residual noise, whereas our MC ADL-MVDR system resolves this issue.

\begin{figure}[t]
  \centering
  \includegraphics[scale = 0.65]{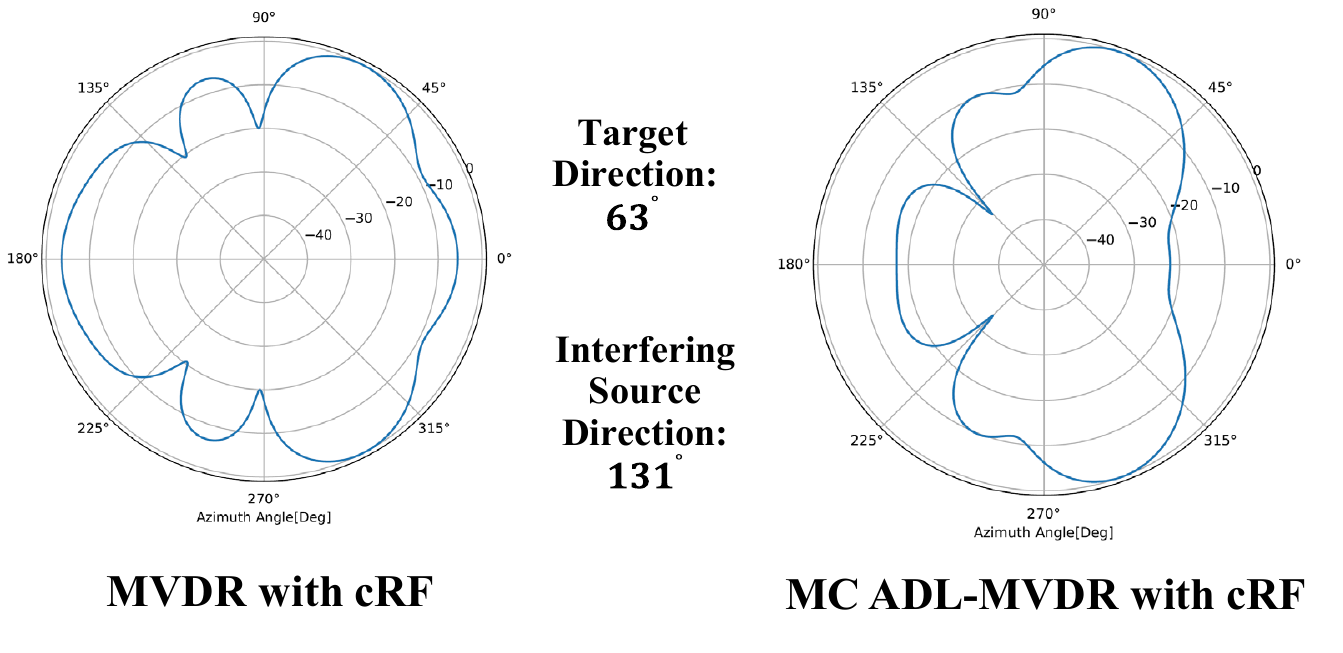}
  \caption{(Color Online). Beam pattern examples for conventional neural mask-based MVDR system and our proposed MC ADL-MVDR system. $\text{RT60}=0.31$ s. Target source at $63^{\circ}$, interfering source at $131^{\circ}$.}
  \label{fig:sample_beampattern}
\end{figure}

An example comparison of the beam patterns between conventional neural mask-based MVDR system and our proposed MC ADL-MVDR system is provided in Fig. \ref{fig:sample_beampattern}. It represents the case of a 2-speaker mixture, with the target and interfering sources at directions of $63^{\circ}$ and $131^{\circ}$, respectively. It is obvious that our proposed MC ADL-MVDR system can better capture the target source information with a sharper main lobe at the corresponding target direction. The frequency for these beam pattern plots is set to 968 Hz. We pick the representative time index for MC ADL-MVDR system in order to visualize its time-varying beamforming weights.

\textbf{MF ADL-MVDR vs. NNs}: The simulation results of the MF ADL-MVDR system are provided in the second block of Table \ref{tab:1}. By comparing the performance between MF ADL-MVDR system and the purely NN system with cRF, we observe that the MF ADL-MVDR system can lead to moderate improvements in all objective metrics (i.e., PESQ: 3.14 vs. 3.10, Si-SNR: 12.60 dB vs. 12.50 dB, SDR: 13.17 dB vs. 13.01 dB, and STOI: 89.5\% vs. 89.2\%) when only one channel is available to the ADL-MVDR module. We infer that the limited improvement here compared to MC ADL-MVDR system is due to the loss of spatial information. In the meantime, the proposed MF ADL-MVDR system achieves huge improvement on ASR accuracy, which is about 11.3\% better in WER than NN with cRF (i.e., 19.57\% vs. 22.07\%). This implies that our proposed MF ADL-MVDR system can greatly reduce the nonlinear distortion introduced by conventional purely NN systems. The purely NN systems (i.e., without MVDR) usually come with a relatively higher degree of nonlinear distortion on separated speech as they only focus on noise reduction \cite{wang2020complex,ochiai2017unified}. As reflected in Fig. \ref{fig:sample_spec}, the highlighted spectral `black holes' (i.e., zero or close to zero values on spectrogram) is an example of the purely NN systems where too much speech information is removed when suppressing the noise, where modern ASR systems are known to be sensitive to this type of distortion \cite{tawara2018adversarial, wang2020complex}. Although not provided with spatial information for the beamforming module, the proposed MF ADL-MVDR system drastically outperforms the purely NN systems in terms of ASR accuracy while achieving better objective scores.

\textbf{MCMF ADL-MVDR vs. MC ADL-MVDR}: The results of the best performing MCMF ADL-MVDR system (9-channel 3-frame) are showed in the last block of Table \ref{tab:1}. Compared to its MC ADL-MVDR peer, we notice that the MCMF ADL-MVDR system can further improve the objective scores (i.e., PESQ: 3.46 vs. 3.42, Si-SNR: 15.43 dB vs. 14.80 dB, SDR: 16.03 dB vs. 15.45 dB, and STOI: 93.7\% vs. 93.3\%), suggesting that MCMF ADL-MVDR system can even better remove the residual noise and generate more intelligible speech signals than MC ADL-MVDR systems. Slightly better performance is observed in terms of ASR accuracy (i.e., 12.31\% vs. 12.73\%) when compared to MC ADL-MVDR system. Results here suggest that on top of the spatial information, incorporating additional MF information is beneficial.

\subsection{Ablation Study on MCMF ADL-MVDR Systems}
\label{subsec:MCMF ADL-MVDR Systems}

\begin{table}[t]
\centering
\caption{Evaluation results for the MCMF ADL-MVDR systems and MC ADL-MVDR systems.}
\label{tab:2}
\scalebox{0.9}{
\begin{tabular}{c|c|c|c|c|c}
\hline \hline
\# of Channels & MF Size & PESQ & Si-SNR (dB) & SDR (dB) & WER (\%) \\ \hline
3 & 3 & \textbf{3.40} & \textbf{14.97} & \textbf{15.50} & \textbf{12.74} \\
3 & 1 & 3.29 & 13.85 & 14.43 & 14.46 \\ \hline
7 & 3 & \textbf{3.43} & \textbf{15.31} & \textbf{15.91} & \textbf{12.84}\\
7 & 1 & 3.34 & 14.14 & 14.80 & 13.67 \\ \hline
9 & 3 & \textbf{3.46} & \textbf{15.43} & \textbf{16.03} & \textbf{12.31} \\
9 & 1 & 3.39 & 14.67 & 15.15 & 13.17 \\ \hline
15 & 2 & \textbf{3.42} & \textbf{15.16} & \textbf{15.73} & 12.89 \\
15 & 1 & \textbf{3.42} & 14.80 & 15.45 & \textbf{12.73} \\ \hline \hline
\end{tabular}}
\end{table}

As shown in Table \ref{tab:2}, we provide comparison results between several MCMF ADL-MVDR systems and MC ADL-MVDR systems (i.e., MF size = 1, current frame). Comparing the performance between MCMF and MC ADL-MVDR systems, we find that the inclusion of additional MF information can often lead to improved performance in both objective scores as well as the ASR accuracy. For example in 3-channel scenario, the PESQ, Si-SNR and SDR scores are 3.40 vs. 3.29, 14.97 dB vs. 13.85 dB, 15.50 dB vs. 14.43 dB between MCMF and MC ADL-MVDR systems. The WER for MCMF ADL-MVDR system is roughly 12\% better than MC ADL-MVDR system (i.e., 12.74\% vs. 14.46\%) in 3-channel scenario.

Additionally, we observe that the gains from the additional MF information for MCMF ADL-MVDR systems become smaller as the number of available microphone channels increases. For instance, in 3-channel scenario, the MCMF ADL-MVDR system achieves relative improvement of 3.3\% on PESQ scores (i.e. 3.40 vs. 3.29) and 12\% on ASR accuracy. Whereas in 7-channel scenario, the relative improvement for MCMF ADL-MVDR system is only around 2.7\% on PESQ scores (i.e., 3.43 vs. 3.34) and about 6\% in terms of WER (i.e., 12.84\% vs. 13.67\%). This pattern suggests that the MF information becomes less important when more spatial information is available. When all 15 channels are provided, similar trend holds such that including additional MF information can further improve the objective scores (e.g., Si-SNR: 15.16 dB vs. 14.80 dB), while achieving similarly high ASR accuracy (i.e., 12.89\% vs. 12.73\%).

We also want to point out that the feature space (i.e., size of the covariance matrix) is increasing exponentially with the number of microphone channels and the MF size. Therefore, when the number of available microphone channels is small, additional MF information (i.e., MCMF ADL-MVDR) may help to enhance the performance. But this could be redundant when more spatial information is available and a large feature size may hinder the learning process of NNs.

\subsection{Ablation Study on cRF Sizes}
\label{subsec:cRF sizes}

\begin{table}[t]
\centering
\caption{Effects of the cRF sizes on the performance of Purely NNs and MC/MF ADL-MVDR systems. The cRF size is represented by its time (T.) and frequency (F.) ranges, where 0, negative and positive numbers indicate the current, previous/lower and future/higher time frame or frequency bin, respectively.}
\label{tab:3}
\scalebox{0.95}{
\begin{tabular}{llllll}
\hline \hline
\multicolumn{1}{l|}{T. Range} & \multicolumn{1}{l|}{F. Range} & \multicolumn{1}{l|}{PESQ} & \multicolumn{1}{l|}{Si-SNR (dB)} & \multicolumn{1}{l|}{SDR (dB)} & WER (\%) \\ \hline
\multicolumn{6}{c}{Purely NN Systems} \\ \hline
\multicolumn{1}{c|}{0} & \multicolumn{1}{c|}{0} & \multicolumn{1}{c|}{3.07} & \multicolumn{1}{c|}{12.23} & \multicolumn{1}{c|}{12.73} & 22.49 \\
\multicolumn{1}{c|}{[-1,1]} & \multicolumn{1}{c|}{[-1,1]} & \multicolumn{1}{c|}{\textbf{3.10}} & \multicolumn{1}{c|}{\textbf{12.50}} & \multicolumn{1}{c|}{\textbf{13.01}} & 22.07\\
\multicolumn{1}{c|}{[-2,2]} & \multicolumn{1}{c|}{[-2,2]} & \multicolumn{1}{c|}{3.09} & \multicolumn{1}{c|}{12.45} & \multicolumn{1}{c|}{12.97} & \textbf{21.80} \\
\multicolumn{1}{c|}{[-3,3]} & \multicolumn{1}{c|}{[-3,3]} & \multicolumn{1}{c|}{3.09} & \multicolumn{1}{c|}{12.44} & \multicolumn{1}{c|}{12.96} & 22.15 \\ \hline
\multicolumn{6}{c}{MC ADL-MVDR Systems} \\ \hline
\multicolumn{1}{c|}{0} & \multicolumn{1}{c|}{0} & \multicolumn{1}{c|}{2.83} & \multicolumn{1}{c|}{11.51} & \multicolumn{1}{c|}{11.98} & 23.73 \\
\multicolumn{1}{c|}{0} & \multicolumn{1}{c|}{[-1,1]} & \multicolumn{1}{c|}{3.21} & \multicolumn{1}{c|}{13.08} & \multicolumn{1}{c|}{13.77} & 17.87 \\
\multicolumn{1}{c|}{[-2,0]} & \multicolumn{1}{c|}{[-1,1]} & \multicolumn{1}{c|}{3.40} & \multicolumn{1}{c|}{14.51} & \multicolumn{1}{c|}{15.09} & 12.88 \\
\multicolumn{1}{c|}{[-1,1]} & \multicolumn{1}{c|}{0} & \multicolumn{1}{c|}{3.37} & \multicolumn{1}{c|}{14.31} & \multicolumn{1}{c|}{14.93} & 13.52 \\ 
\multicolumn{1}{c|}{[-1,1]} & \multicolumn{1}{c|}{[-1,1]} & \multicolumn{1}{c|}{\textbf{3.42}} & \multicolumn{1}{c|}{\textbf{14.80}} & \multicolumn{1}{c|}{\textbf{15.45}} & \textbf{12.73} \\ \hline
\multicolumn{6}{c}{MF ADL-MVDR Systems} \\ \hline
\multicolumn{1}{c|}{0} & \multicolumn{1}{c|}{0} & \multicolumn{1}{c|}{2.95} & \multicolumn{1}{c|}{11.75} & \multicolumn{1}{c|}{12.27} & 21.97 \\
\multicolumn{1}{c|}{0} & \multicolumn{1}{c|}{[-1,1]} & \multicolumn{1}{c|}{3.12} & \multicolumn{1}{c|}{12.26} & \multicolumn{1}{c|}{12.81} & 21.08 \\
\multicolumn{1}{c|}{[-1,0]} & \multicolumn{1}{c|}{0} & \multicolumn{1}{c|}{3.13} & \multicolumn{1}{c|}{12.44} & \multicolumn{1}{c|}{12.96} & 20.63 \\
\multicolumn{1}{c|}{[-2,0]} & \multicolumn{1}{c|}{0} & \multicolumn{1}{c|}{3.13} & \multicolumn{1}{c|}{12.44} & \multicolumn{1}{c|}{12.95} & 21.11 \\
\multicolumn{1}{c|}{[-2,0]} & \multicolumn{1}{c|}{[-1,1]} & \multicolumn{1}{c|}{3.11} & \multicolumn{1}{c|}{12.35} & \multicolumn{1}{c|}{12.90} & 21.41 \\
\multicolumn{1}{c|}{[-1,1]} & \multicolumn{1}{c|}{0} & \multicolumn{1}{c|}{\textbf{3.16}} & \multicolumn{1}{c|}{12.55} & \multicolumn{1}{c|}{13.04} & 20.40 \\ 
\multicolumn{1}{c|}{[-1,1]} & \multicolumn{1}{c|}{[-1,1]} & \multicolumn{1}{c|}{3.14} & \multicolumn{1}{c|}{\textbf{12.60}} & \multicolumn{1}{c|}{\textbf{13.17}} & \textbf{19.57} \\
\multicolumn{1}{c|}{[-2,2]} & \multicolumn{1}{c|}{[-2,2]} & \multicolumn{1}{c|}{3.15} & \multicolumn{1}{c|}{12.58} & \multicolumn{1}{c|}{13.12} & 20.42 \\
\hline \hline
\end{tabular}}
\end{table}

The results for different sizes of the cRF are presented in Table \ref{tab:3}. The MF size is fixed at five (i.e., from two previous frames to two future frames) for all MF ADL-MVDR systems. We find that a $1\times1$ cRF (i.e., a cRM) results in the worst performance (e.g., WER: 22.49\%) for purely NN systems when compared its peers with larger cRF sizes. The NN with a $5\times5$ cRF (i.e., the 3rd purely NN system) leads to the best performance in WER (i.e., 21.80\%) and a $3\times3$ cRF can achieve the best objective scores (i.e., PESQ: 3.10, Si-SNR: 12.50 dB, SDR: 13.01 dB). In general, we find that NNs with cRF of sizes $3\times3$, $5\times5$ and $7\times7$ yield with similar performance in objective metrics and ASR accuracy, while the cRM alone is not sufficient to achieve the optimal performance. Larger cRF size could lead to slightly better performance for purely NNs but there is also a trade-off on the performance and run time efficiency of the system. 

Similar patterns can be found in MC ADL-MVDR systems, where the cRM alone (i.e., the 1st MC ADL-MVDR system) is not leading to satisfactory performance (e.g., WER: 23.73\%) and size of the cRF could be even more important. By comparing the systems with cRF of sizes $1\times1$ and $1\times3$ (i.e., the 1st and 2nd MC ADL-MVDR systems), we find that including nearby frequency information would help improve the system's performance (e.g., WER: 23.73\% vs. 17.87\%). Whereas substantial improvements (e.g., WER: 23.73\% vs. 13.52\%) can be achieved by introducing the nearby temporal information to the cRF (i.e., the 1st and 4th MC ADL-MVDR systems), which suggests that temporal information could be more important than nearby frequency information for our proposed ADL-MVDR systems. Meanwhile, we also find that including future frame information in cRF could help improve the system's performance. For example, by comparing the MC ADL-MVDR systems with cRF that contains information from two previous frames and the other one with temporal information from one previous frame to one future frame (i.e., the 3rd and 5th MC ADL-MVDR systems), slight improvements can be observed in both ASR accuracy (i.e., 12.73\% vs. 12.88\%) and objective scores (i.e., PESQ: 3.42 vs. 3.40, Si-SNR: 14.80 dB vs. 14.51 dB and SDR: 15.45 dB vs. 15.09 dB).

Several cRF setups for MF ADL-MVDR systems are also included and their results are generally consistent with those for MC ADL-MVDR and purely NN systems. Specifically, a $1\times1$ cRF (i.e., a cRM) does not perform well on MF ADL-MVDR system either (i.e., lowest ASR accuracy and objective scores). The inclusion of future frame information in cRF is crucial (e.g., comparing the 4th and 6th MF ADL-MVDR systems, WER: 21.11\% vs. 20.40\%) and that the nearby frequency information could also improve the ASR performance slightly while achieving similar objective scores (i.e., comparing the 6th and 7th MF ADL-MVDR systems, PESQ: 3.16 vs. 3.14, Si-SNR: 12.55 dB vs. 12.60 dB, SDR: 13.04 dB vs. 13.17 dB, and WER: 20.40\% vs. 19.57\%). Increasing the cRF size from $3\times3$ to $5\times5$ (i.e., the last two MF ADL-MVDR systems) does not improve the system's performance and even result in slightly poorer performance (except for PESQ scores), which indicates that adopting a very large size cRF may not be necessary or beneficial. In general, incorporating future T-F pixels to the cRF can improve the systems performance but in a limited scale, which could be caused by the non-causality of the complex filter estimator that already uses some future information.

\begin{table}[t]
\centering
\caption{Effects of MF sizes on the objective performance and ASR accuracy of MF ADL-MVDR systems. $t$ represents the current frame.}
\label{tab:4}
\scalebox{0.92}{
\begin{tabular}{llllll}
\hline \hline
\multicolumn{6}{c}{MF ADL-MVDR Systems} \\ \hline
\multicolumn{1}{l|}{MF Size} & \multicolumn{1}{l|}{MF Range} & \multicolumn{1}{l|}{PESQ} & \multicolumn{1}{l|}{Si-SNR (dB)} & \multicolumn{1}{l|}{SDR (dB)} & WER (\%) \\ \hline
\multicolumn{1}{c|}{2} & \multicolumn{1}{c|}{[t-1,t]} & \multicolumn{1}{c|}{3.02} & \multicolumn{1}{c|}{12.06} & \multicolumn{1}{c|}{12.58} & 20.50 \\
\multicolumn{1}{c|}{3} & \multicolumn{1}{c|}{[t-2,t]} & \multicolumn{1}{c|}{3.12} & \multicolumn{1}{c|}{12.39} & \multicolumn{1}{c|}{12.95} & 20.87 \\
\multicolumn{1}{c|}{3} & \multicolumn{1}{c|}{[t-1,t+1]} & \multicolumn{1}{c|}{3.09} & \multicolumn{1}{c|}{12.13} & \multicolumn{1}{c|}{12.71} & 20.27 \\
\multicolumn{1}{c|}{5} & \multicolumn{1}{c|}{[t-2,t+2]} & \multicolumn{1}{c|}{3.14} & \multicolumn{1}{c|}{12.60} & \multicolumn{1}{c|}{13.17} & 19.57 \\
\multicolumn{1}{c|}{7} & \multicolumn{1}{c|}{[t-3,t+3]} & \multicolumn{1}{c|}{\textbf{3.18}} & \multicolumn{1}{c|}{\textbf{12.89}} & \multicolumn{1}{c|}{\textbf{13.42}} & \textbf{19.45} \\
\multicolumn{1}{c|}{9} & \multicolumn{1}{c|}{[t-4,t+4]} & \multicolumn{1}{c|}{3.17} & \multicolumn{1}{c|}{12.76} & \multicolumn{1}{c|}{13.27} & 19.52 \\
\hline \hline
\end{tabular}}
\end{table}

\subsection{Ablation Study on MF Sizes}
\label{subsec:Multi-frame Size on Single-channel System}

\begin{table*}[t]
\centering
\caption{Model size, memory usage and running speed information of some evaluated systems.}
\label{tab:5}
\scalebox{0.98}{
\begin{tabular}{l|c|c|c|c|c|c|c|c}
\hline \hline
Systems & Model Size & \multicolumn{1}{c|}{\begin{tabular}[c]{@{}c@{}}Training Mem.\\ (MB)\end{tabular}} & \multicolumn{1}{c|}{\begin{tabular}[c]{@{}c@{}}Inference Mem.\\ (MB)\end{tabular}} & \multicolumn{1}{c|}{\begin{tabular}[c]{@{}c@{}}Training Speed\\ (ms)\end{tabular}} & \multicolumn{1}{c|}{\begin{tabular}[c]{@{}c@{}}Inference Speed\\ (ms)\end{tabular}} & PESQ & \multicolumn{1}{c|}{\begin{tabular}[c]{@{}c@{}}Si-SNR\\ (dB)\end{tabular}}  & \multicolumn{1}{c}{\begin{tabular}[c]{@{}c@{}}WER\\ (\%)\end{tabular}} \\ \hline
\multicolumn{9}{c}{Purely NNs and our proposed MF ADL-MVDR System} \\ \hline
NN with cRM & $\textbf{16.40 M}$ & $\textbf{176.4}$ & $\textbf{67.9}$ & $\textbf{53.7}$ & $\textbf{13.5}$ & 3.07 & 12.23 & 22.49 \\
NN with cRM (large) & 31.62 M & 335.0 & 126.3 & 54.5 & 14.1 & 3.07 & 12.26 & 22.06 \\
NN with cRF & 17.46 M & 193.3 & 70.3 & 66.4 & 15.7 & 3.10 & 12.50 & 22.07 \\ \hline
MF ADL-MVDR & 18.99 M & 288.5 & 78.7 & 108.4 & 20.1 & $\textbf{3.14}$ & $\textbf{12.60}$ & $\textbf{19.57}$ \\ \hline
\multicolumn{9}{c}{Conventional neural mask-based MVDR and our proposed MC/MCMF ADL-MVDR systems} \\ \hline
MVDR with cRF - recursive \cite{higuchi2016robust} & $\textbf{18.32 M}$ & 281.2 & 74.1 & 174.8 & 31.3 & 2.90 & 10.12 & 16.74 \\
MVDR with cRF & $\textbf{18.32 M}$ & 282.8 & 74.5 & $\textbf{83.1}$ & $\textbf{16.4}$ & 2.92 & 11.31 & 15.91 \\
MVDR with cRF (large) & 35.84 M & 547.1 & 141.5 & 83.7 & 17.0 & 2.94 & 11.38 & 15.75 \\
Multi-tap MVDR with cRF & $\textbf{18.32 M}$ & 284.4 & 74.6 & 98.4 & 20.8 & 3.08 &  12.66 & 13.52 \\ \hline
MC ADL-MVDR (compact) & $\textbf{18.32 M}$ & $\textbf{280.5}$ & $\textbf{74.0}$ & 131.9 & 27.0 & 3.37 & 14.36 & 13.01 \\
MC ADL-MVDR & 23.47 M & 359.9 & 93.6 & 222.2 & 38.9 &  3.42 & 14.80  &  12.73 \\
MCMF ADL-MVDR & 27.01 M & 416.2 & 106.4 & 302.6 & 55.3 & $\textbf{3.46}$ & $\textbf{15.43}$ & $\textbf{12.31}$ \\
\hline \hline
\end{tabular}
}
\end{table*}

The cRF sizes for all MF ADL-MVDR systems presented in Table \ref{tab:4} are fixed at $3\times3$ (i.e., $\pm1$ nearby frames and frequency bins) in order to investigate the influence of different MF sizes. As shown in Table \ref{tab:4}, we include six different setups where the first two system setups represent the conditions when only information from previous frames is available, and the last four MF conditions further include information from future frames. By comparing the first two MF ADL-MVDR system in Table \ref{tab:4}, we find that the inclusion of additional previous frame (i.e., t-2) could help improve the objective scores (e.g., Si-SNR: 12.39 dB vs. 12.06 dB) while achieving similar WER performance (i.e., 20.87\% vs. 20.50\%). Then, by comparing the 1st and 3rd MF ADL-MVDR systems (i.e., [t-1,t] and [t-1,t+1]), we observe improvements in both the objective metrics (e.g., PESQ: 3.02 vs. 3.09) and ASR accuracy (i.e., 20.50\% vs. 20.27\%), indicating that it is beneficial to include future frames in the MF information. By increasing the MF range (e.g., the 3rd and 4th MF ADL-MVDR systems in Table \ref{tab:4}), further improvements can be obtained (e.g., PESQ: 3.09 vs. 3.14 and WER: 20.27\% vs. 19.57\%). We also find that further expanding the MF size leads to even better performance (i.e., the 5th and 4th MF ADL-MVDR systems), where the system achieves better objective scores (e.g., PESQ: 3.18 vs. 3.14) as well as improved ASR accuracy (i.e., WER: 19.45\% vs. 19.57\%). However, when the MF size continually increases from seven to nine (i.e., the last two setups in Table \ref{tab:4}), no further improvements are observed (e.g., PESQ: 3.17 vs. 3.18 and WER: 19.52\% vs. 19.45\%). Better performance can be achieved by increasing the MF size, however, additional information from future frames did not make significant contributions to the performance, which is likely caused by the non-causality of the current front-end audio encoding network.

\subsection{Ablation Study on Model Statistics}
\label{subsec:ablation study model size}

Comparison results of the statistics in the systems' training and inference stages are reported in Table \ref{tab:5}. A single Nvidia Tesla V100 GPU is used to measure the statistics, where we set the batch size to 1 for evaluation in both training and inference stage. The model size is first reported by the total number of parameters in a model, memory usage information\footnote{pytorch\_memlab, \href{https://github.com/Stonesjtu/pytorch_memlab}{https://github.com/Stonesjtu/pytorch\_memlab}} (training and inference stages) and running speed (average running time for processing 1 s of audio input) are also included. The systems' performance on PESQ, Si-SNR and WER are provided for comparison.

The purely NN with cRM has the smallest model size (i.e., 16.40 M) since it does not come with a cRF estimator or RNN-based beamforming module. The proposed MF ADL-MVDR with cRF has slightly larger model size compared to NN with cRF (i.e., 18.99 M vs. 17.46 M), but with similar inference memory (i.e., 78.7 MB vs. 70.3 MB), inference speed (20.1 ms vs. 15.7 ms) while better performance in speech quality (e.g., PESQ: 3.14 vs. 3.10) and ASR accuracy (i.e., 19.57\% vs. 22.07\%).

The proposed MC ADL-MVDR systems utilize additional spatial information (i.e., microphone channels) in the GRU-Nets that leads to increased model size compared to conventional neural mask-based MVDR systems (e.g., 23.47 M vs. 18.32 M) that is based on mathematical derivation. It obtains much better performance compared to multi-tap MVDR system on the separated speech quality (i.e., PESQ: 3.42 vs. 3.08, Si-SNR: 14.80 dB vs. 12.66 dB) and WER (i.e., 12.73\% vs. 13.52\%), at the cost of additional inference memory (i.e., 93.6 MB vs. 74.6 MB) and inference time (i.e., 38.9 ms vs. 20.8 ms). When more computation resources are available, the MCMF ADL-MVDR system leads to the best performance (i.e., PESQ; 3.46, Si-SNR: 15.43 dB and WER 12.31\%) with a model size of 27.01 M and slightly increased inference memory and speed compared to MC ADL-MVDR (i.e., 106.4 MB and 55.3 ms, respectively).

To investigate whether the performance gain is introduced by increased number of network parameters, we further include three models, (1) a lager purely NN with cRM, denoted with NN with cRM (large), (2) a larger neural mask-based MVDR system with cRF, denoted as MVDR with cRF (large), and (3) a compact version of the MC ADL-MVDR system, denoted as MC ADL-MVDR (compact). For the two enlarged models, we increase the embedding dimension of the audio encoding block from 256 to 512, with everything else remained unchanged. Meanwhile, for MC ADL-MVDR (compact), we reduce the hidden feature space of the GRU-Nets (i.e., 250 and 50 for $\mathbf{GRU{\text-}Net}_{\boldsymbol{v}}$, 250 for $\mathbf{GRU{\text-}Net}_{\boldsymbol{\mathrm{NN}}}$) and the cRF region (i.e., 3 $\times$ 1 cRF, only using nearby temporal information).

As showed in Table \ref{tab:5}, simply increasing the network size does not lead to major improvements on the systems' performance. For purely NNs with cRM, we have PESQ: 3.07 vs. 3.07, Si-SNR: 12.26 dB vs. 12.23 dB and WER: 22.06\% vs. 22.49\% between the large and original models. On the other hand, for MVDR systems with cRF, increasing the model size leads to 0.02 and 0.07 dB increments in PESQ and Si-SNR, respectively. Even with a reduced model size in MC ADL-MVDR (compact), it still achieves relative 15\%, 27\% and 18\% gains on PESQ, Si-SNR, and WER compared to MVDR with cRF. For fair comparison on model statistics, the proposed ADL-MVDR systems should be compared with the recursive MVDR system, as they estimate time-varying beamforming weights and can be adapted for online processing with streaming input. Note that conventional chunk-level MVDR systems cannot be directly used for this purpose as they require a chunk of audio input to derive the beamforming weights. When compared to MVDR system with recursive updating rules, the MC ADL-MVDR (compact) achieves faster inference speed (i.e., 27.0 ms vs. 31.3 ms), lower inference memory (i.e., 74.0 MB vs. 74.1 MB) while gaining substantial gains on speech quality (e.g., PESQ: 3.37 vs. 2.90) and ASR accuracy (13.01\% vs. 16.74\%).

Results here indicate that improvements are limited for purely NNs and conventional neural mask-based MVDR systems with increased model size, and our proposed MC ADL-MVDR system still achieves substantial improvements than its peers with the same model size. We infer that the additional inference time and memory cost in our proposed ADL-MVDR systems compared to purely NNs and conventional neural mask-based MVDR systems are coming from the RNN-based networks in the ADL-MVDR module. Moreover, when compared to conventional neural mask-based MVDR system with recursive updating rules, the proposed ADL-MVDR systems consume similar computing resources and time while delivering much better performance on speech quality as well as the ASR accuracy.

\section{Conclusions and Future Work}
\label{sec:Conclusions}

In this work, we proposed an ADL-MVDR framework that can be configured and applied for multi-channel, multi-frame, and multi-channel multi-frame target speech separation tasks. Our proposed ADL-MVDR systems have achieved the best performance in terms of objective speech quality and intelligibility, as well as ASR accuracy in both the multi-channel and multi-frame scenarios among its peers. The multi-channel multi-frame ADL-MVDR system can achieve even better performance by fully exploring spatio-temporal cross correlations. Leveraging on RNN-predicted filtering weights, the proposed ADL-MVDR system also bypasses the numerical instability issue that occurs in conventional neural mask-based MVDR systems during joint training with neural networks. Additionally, the proposed ADL-MVDR systems keep the residual noise at a minimum level (reflected by highest objective scores) while introducing hardly any nonlinear distortions (reflected by lowest WER). However, one limitation of the current ADL-MVDR framework is that it requires additional computation resources compared to conventional neural mask-based MVDR approaches.

Future work of this study may include the following directions. First, we will evaluate this ADL-MVDR framework on purely single-channel speech separation tasks. Second, the complex ratio filter and multi-frame sizes are determined empirically in this study, we will further explore approaches to control these sizes adaptively. Third, we will adapt this ADL-MVDR framework for joint separation and dereverberation tasks. Fourth, we will explore a more generalized solution of neural beamformer that is independent of microphone geometries. Lastly, we will further explore on different network structures to provide a more efficient model, for instance, we could try replacing the RNN-based networks with feed-forward mechanisms for improved efficiency.

\ifCLASSOPTIONcaptionsoff
  \newpage
\fi

\bibliographystyle{IEEEtran}
\bibliography{IEEEabrv,refs}

\end{document}